\documentclass[a4paper,twocolumn,showpacs,superscriptaddress,floatfix,prd,nofootinbib]{revtex4-1}
\usepackage{latexsym}
\usepackage{graphicx,epsf,epsfig}
\usepackage{amssymb,amsmath}
\usepackage{bm}
\usepackage[usenames]{color}
\usepackage{pstricks}
\usepackage{rotating}
\def\be{\begin{equation}}
\def\ee{\end{equation}}
\def\beq{\begin{eqnarray}}
\def\eeq{\end{eqnarray}}

\begin{document}

\title{On the validity of the adiabatic approximation in compact binary inspirals}

\author{Andrea Maselli}
\affiliation{Dipartimento di Fisica, ``Sapienza'' Universit\`a di Roma
  \& Sezione INFN Roma1, Rome, Italy
}
\author{Leonardo Gualtieri}
\affiliation{Dipartimento di Fisica, ``Sapienza'' Universit\`a di Roma
  \& Sezione INFN Roma1, Rome, Italy
}

\author{Francesco Pannarale}
\affiliation{
Max-Planck-Institut f{\"u}r Gravitationsphysik, Albert Einstein
Institut, Potsdam, Germany
}

\author{Valeria Ferrari}
\affiliation{Dipartimento di Fisica, ``Sapienza'' Universit\`a di Roma
  \& Sezione INFN Roma1, Rome, Italy
}
\begin{abstract} 
Using a semi-analytical approach recently developed to model the tidal
deformations of neutron stars in inspiralling compact binaries, we study
the dynamical evolution of the tidal tensor, which we explicitly derive at
second post-Newtonian order, and of the quadrupole tensor. Since we do not
assume {\it a priori} that the quadrupole tensor is proportional to the tidal
tensor, i.e. the so called ``adiabatic approximation'', our approach enables
us to establish to which extent such approximation is reliable. We find that
the ratio between the quadrupole and tidal tensors (i.e., the Love number)
increases as the inspiral progresses, but this phenomenon only marginally
affects the emitted gravitational waveform. We estimate the frequency range in
which the tidal component of the gravitational signal is well described using
the stationary phase approximation at next-to-leading post-Newtonian order,
comparing different contributions to the tidal phase. We also derive a
semi-analytical expression for the Love number, which reproduces within a few
percentage points the results obtained so far by numerical
integrations of the relativistic equations of stellar perturbations.
\end{abstract}

\maketitle

\section{Introduction}\label{intro}
Coalescing binary systems of neutron stars (NS) and/or black holes (BH) are
among the most interesting sources of gravitational waves (GWs) to be detected
by advanced Virgo and LIGO \cite{virgoligo}. One of the key features of the
coalescence is the NS tidal deformation, which provides precious information on
the NS equation of state (EOS) \cite{FH08,HLLR10,DN09,DN10,PROR11}. For this
reason, theoretical and numerical studies have been recently performed to model
the effect of tidal deformations on the emitted GW signal, and to extract its
contribution from a detected signal \cite{FH08, RMSUCF09, BDGNR10, HLLR10, DN10,
  PROR11, VFH11, LKSBF12, BNTB12}. These studies are based on fully
general-relativistic numerical simulations and on (semi-)analytical approaches.

In current analytical approaches, the tidal deformation properties of NSs are
encoded in a set of numbers, the {\it Love numbers}
\cite{FH08,H08,HLLR10,DN09,BP09,DN10,PROR11}, which relate the mass multipole
moments of the star to the (external) tidal field multipole moments. In the
so-called ``adiabatic approximation'', at the lowest multipole order, the
evolution of a star in response to an external quadrupolar tidal field
$C_{ij}$,
is governed by the equation
\begin{equation}
\label{ad}
Q_{ij}=-\frac{2}{3}k_2R_\text{NS}^5C_{ij}\,,
\end{equation}
where $Q_{ij}$ is the star quadrupole moment, $R_\text{NS}$ is its radius at
isolation, and $k_2$ is the (dimensionless) $l=2$ apsidal constant, also dubbed
{\it second tidal Love number}. For non-rotating relativistic stars, $k_2$ is
usually computed by solving the linear $l=2$ static, even-parity perturbations
of Tolman-Oppenheimer-Volkoff solutions \cite{TC67}, requiring regularity and
continuity for the metric perturbation and its first derivative
\cite{H08,DN09,BP09}. Eq.~(\ref{ad}) and its higher multipole order versions
have been employed to determine the effect of tidal deformations on the orbital
motion, and on the GW signal emitted by NS-NS and BH-NS binary systems
\cite{HLLR10,DN10,VFH11,PROR11}. In these studies, the relation in
Eq.~(\ref{ad}) --- or an equivalent assumption \cite{DN10} --- is assumed as a
starting point.

The domain of validity of (\ref{ad}) is briefly discussed in \cite{FH08}, using
previous results of Lai. In \cite{L94} a NS tidally interacting with a
companion, is described as a forced oscillator in a Newtonian framework;
the energy absorbed by the oscillator is the sum of an ``instantaneous''
term, proportional to the forcing (tidal) field, and a term associated with the
stellar oscillations. Since the former is much larger than the latter, it is
argued that the adiabatic approximation (\ref{ad}) holds \cite{FH08}. A forced oscillator
model is also used in \cite{HLLR10}, providing further evidence of the accuracy
of the adiabatic approximation. It should be stressed, however, that only a
consistent dynamical study of the stellar deformation during the inspiral may
assess the validity of the adiabatic approximation. This is the approach we follow
in this paper.

In a recent work \cite{FGM11}, some of us developed a semi-analytical
description of tidal deformations of NSs in inspiralling BH-NS and NS-NS
binaries. This model combines the post-Newtonian (PN) approach (see, for
instance, \cite{B06,BAF06,BAF11}), which accurately describes the orbital motion
before the onset of mass-shedding, and the {\it affine model}, which allows for
a description of stellar deformations due to an external quadrupolar tidal field
generated by the companion
\cite{CL85,LM85,WL00,CFS06,FGP09,FGP10}. We computed the tidal tensor associated
with the PN metric of a two-body system, defined in terms of the PN Riemann
tensor and of the NS local tetrad
\begin{equation}
  C_{ij}=R_{\alpha\beta\gamma\delta}
e^{\alpha}_{(0)}e^{\beta}_{(i)}e^{\gamma}_{(0)}e^{\delta}_{(j)}\,,
\label{defC}
\end{equation}
up to $1.5$PN order. This tensor appears as a source in the dynamical equations
describing the stellar deformations. Note that the use of a two-body PN metric
explicitly yields self-interaction terms in the tidal tensor. We validated our
post-Newtonian-affine (PNA) approach by comparing the results obtained for BH-NS
binaries to the outcome of fully general-relativistic numerical simulations. In
addition, we computed the Love number $k_2$ using Eq.~(\ref{ad}) at large
separations, showing that our results were in good agreement with the analytical
results of \cite{H08}.

In this paper, we extend the PNA approach further:
\begin{itemize}
\item We compute the tidal tensor (\ref{defC}) up to $2$PN order.  This new
  result may allow one to construct more accurate models of gravitational
  waveforms emitted by inspiralling compact binaries, which could be employed to
  extract information on the Love number and on the underlying equation of
  state, during post-processing. Indeed, the $2$PN terms may be used to correct
  the binary binding energy and the GW-flux and to improve the description of
  the phase evolution.
\item We use the $2$PN tidal tensor to assess the range of validity of the
  adiabatic approximation. We do so by solving the dynamical equations for the
  orbital motion and the stellar deformations, by computing the quadrupole and
  tidal tensors, and by determining $k_2$ from
\begin{equation}
\label{k2}
k_2 = -\frac{3Q_{ij}}{2R_\text{NS}^5C_{ij}}\,,
\end{equation}
at different values of the orbital separation. We find that $k_2$ increases in
the late inspiral phase; it should thus be referred to as a {\it Love function}
$k_2(r)$, of which $k_2$ is the asymptotic limit.  As discussed in Section
\ref{sec:love}, we prefer to express $k_2$ as a function of r, rather than as a
function of the gauge invariant frequency, because in this way it is easier to
find an accurate fit for $k_2$.  We then use the stationary phase approximation
\cite{SD91,CF94} to compute the gravitational waveforms with tidal phase effects
included up to $1$PN relative order \cite{VFH11} and model the gravitational
wave phase accordingly.

\item We compute the fitting factors \cite{LOB08} between point-particle
  templates and a model of the gravitational signal that includes tidal effects
  at the best of our knowledge up to $1$PN order, i.e., including the Love
  function $k_2(r)$.  We find that point particle templates satisfy the accuracy
  standards defined in \cite{LOB08}, unless one considers NS-NS binaries with a
  very stiff equation of state, in which case the aforementioned standards are
  marginally violated. Moreover, we show that tidal effects can affect the
  measurement of the total mass and the symmetric mass ratio by at most $3$\%
  and $2$\%, respectively.

\item We derive a simple, semi-analytical expression for the tidal
  Love number $k_2$:
\[
k_2=\frac{15}{4}\frac{\hat{\cal M}^2}{\hat\Pi\, R_\text{NS}^5}\,,
\]
where $\hat{\cal M}$ and $\hat\Pi$ are, respectively, the scalar quadrupole
moment of the star and the integral of the pressure over the stellar volume,
both calculated for the star at isolation.  We check the accuracy of this
formula by comparing its results to those obtained by perturbative approaches
\cite{H08}, finding that they agree within a few percentage points.

\end{itemize}

The plan of the paper is the following. In Sec~\ref{model} we briefly describe
our model. In Sec.~\ref{sec:love} we study the tidal Love number $k_2$ in the
post-Newtonian affine approach. In Sec.~\ref{sec:data} we compute the
gravitational waveform in the stationary phase approximation, comparing the
different tidal contributions to the Fourier phase. In Sec.~\ref{wavecomp} we
compare the waveforms obtained by using different approximations for the tidal
contributions to the Fourier phase. In Sec.~\ref{conclusions} we draw the
conclusions.

\section{The model}\label{model}
In the following we briefly sketch the post-Newtonian-affine model. For further
details, see \cite{FGM11}. The masses of the two compact objects,
inspiralling on quasi-circular orbits, are $m_{1}$ and $m_{2}$, 
$m=m_{1}+m_{2}$ is the total mass, and
$\nu=m_{1}m_{2}/m^{2}$ is the symmetric mass ratio.
We write the  equations  for the
secondary object $m_2$, of radius  $R_\text{NS}$. 
In a NS-NS system, analogous equations hold for the primary object $m_1$.

\subsection{Tidal deformations in the affine model}\label{sec:aff}
To describe the stellar deformation, we use the affine model approach
\cite{CL85,LM85,WL00} (improved in \cite{FGP09,FGP10,FGM11}), which is based on
the assumption that a NS in a binary system preserves an ellipsoidal shape when
it is deformed by the tidal field of the companion. The deformation equations
are written in the principal frame, i.e., the one comoving with the star and
with axes coincident with the principal axes of the ellipsoid. Surfaces of
constant density inside the star form self-similar ellipsoids, and the velocity
of a fluid element is a linear function of the coordinates $x^{i}$ of the
principal frame. Under these assumptions, the infinite degrees of freedom of the
stellar fluid reduce to five dynamical variables and we are left with a set of
ordinary differential equations describing the evolution of the star. The five
variables describing the stellar deformation are the three principal axes of the
ellipsoid\footnote{In our conventions, '1' denotes the direction along the axis
  directed towards the companion, '2' is the direction along the other axis that
  lies in the orbital plane, and '3' indicates the direction of the axis
  orthogonal to the orbital plane.} $a_{i}$ $(i=1,2,3)$ and two angles $\psi$,
$\lambda$ defined as
\begin{equation}\label{NSspin}
  \frac{d\psi}{d\tau}=\Omega\ ,\qquad \frac{d\lambda}{d\tau}=\Lambda\ , 
\end{equation} 
where $\tau$ is the NS proper time, $\Omega$ is the ellipsoid angular velocity
measured in the parallel transported frame associated with the star center of
mass (the ``figure'' velocity in \cite{WL00}), and $\Lambda$ (defined in \cite{WL00}
in terms of the vorticity along the $z-$axis in the corotating frame) describes
the internal fluid motion in the principal frame.

The NS internal dynamics is described in terms of the Lagrangian
\begin{equation}
{\cal L}_\text{I}=T_\text{I}-U-{\cal V}\label{afflag}
\end{equation}
where $T_\text{I}$ is the star kinetic energy, $U$ is the internal energy of the
stellar fluid, and ${\cal V}$ is the star self-gravity.  The relevant integrals
are the scalar quadrupole moment ${\cal M}=1/3\,\int dM_B\sum_i(x_i)^2$, the
pressure integral $\Pi=\int dM_B\,p/\rho$, and the self-gravity potential ${\cal
  V}=-\int dM_B \sum_ix_i\partial_i\Phi$; here $dM_B$ is the baryon mass element, $\Phi$
is the gravitational potential, $p$ is the pressure, and $\rho$ is the baryon
mass density.

 As shown in \cite{CL85,LM85}, under the affine hypothesis the integrals ${\cal
   M}$, $\Pi$, ${\cal V}$ can be expressed in terms of integrals over the
 spherical configuration of the star ($a_i=R_\text{NS}$), and of functions of the
 dynamical variables. In the following, all carets ( $\hat{\ }$ ) denote
 quantities computed on the spherical star. In the spherical configuration, the
 pressure integral and the self-gravity are related by the virial theorem
 \begin{equation}
 \hat{\cal V}=-3\hat\Pi\,.
 \label{virial}
 \end{equation}
 The original affine approach, introduced in a Newtonian framework
 \cite{CL85,LM85,WL00}, was improved and extended in \cite{FGP09,FGP10,FGM11}.
 The spherical configuration of the star is determined by solving the
 relativistic equations of stellar structure, which yield the profile of $\hat
 p(r_s)$, being $r_s$ the radial coordinate in a Schwarzschild frame associated
 to the NS. The gravitational potential $\Phi$ appearing in the definition of
 ${\cal V}$ is replaced by an effective relativistic potential such that the
 virial theorem (\ref{virial}) is satisfied. The equations of motion for the
 internal variables $q_{i}=\{\psi,\lambda,a_{1},a_{2},a_{3}\}$ and their
 conjugate momenta
 $p_{i}=\{p_{\psi},p_{\lambda},p_{a_{1}},p_{a_{2}},p_{a_{3}}\}$ are:
\begin{eqnarray}
  \frac{da_{1}}{dt}&=&\frac{R_\text{NS}}{\gamma(t)}\frac{p_{a_{1}}}{\hat{\mathcal{M}}}\label{da1}\\
  \frac{da_{2}}{dt}&=&\frac{R_\text{NS}}{\gamma(t)}\frac{p_{a_{2}}}{\hat{\mathcal{M}}}\label{da2}\\
  \frac{da_{3}}{dt}&=&\frac{R_\text{NS}}{\gamma(t)}\frac{p_{a_{3}}}{\hat{\mathcal{M}}}\label{da3}
\end{eqnarray}
\begin{eqnarray}
  \frac{dp_{a_{1}}}{dt}&=&\frac{\hat{\mathcal{M}}}{\gamma(t)}\Bigg[\Lambda^{2}+
  \Omega^{2}-2\frac{a_{2}}{a_{1}}\Lambda\Omega+\frac{1}{2}
  \frac{\hat{\cal V}}{\hat{\mathcal{M}}}R_\text{NS}^{3}\tilde{A}_{1}\nonumber\\
  &+&\frac{R_\text{NS}^{2}}{\hat{\mathcal{M}}}\frac{\Pi}{a_{1}^{2}}-c_{xx}\Bigg]a_{1}\label{eqa1}\\
  \frac{dp_{a_{2}}}{dt}&=&\frac{\hat{\mathcal{M}}}{\gamma(t)}\Bigg[\Lambda^{2}
  +\Omega^{2}-2\frac{a_{1}}{a_{2}}\Lambda\Omega+
  \frac{1}{2}\frac{\hat{\cal V}}{\hat{\mathcal{M}}}R_\text{NS}^{3}\tilde{A}_{2}\nonumber\\ 
  &+&\frac{R_\text{NS}^{2}}{\hat{\mathcal{M}}}\frac{\Pi}{a_{2}^{2}}-c_{yy}\Bigg]a_{2}\label{eqa2}\\
  \frac{dp_{a_{3}}}{dt}&=&\frac{\hat{\mathcal{M}}}{\gamma(t)}\left[\frac{1}{2}
    \frac{\hat{\cal V}}{\hat{\mathcal{M}}}R_\text{NS}^{3}\tilde{A}_{3}
    +\frac{R_\text{NS}^{2}}{\hat{\mathcal{M}}}\frac{\Pi}{a_{3}^{2}}-c_{zz}\right]a_{3}\label{eqa3}\\
  \frac{d\lambda}{dt}&=&\frac{\Lambda}{\gamma(t)}\label{dlambda}\\
  \frac{dp_{\lambda}}{dt}&=&\frac{\dot{\mathcal{C}}}{\gamma(t)}=0\\
  \frac{d\psi}{dt}&=&\frac{\Omega}{\gamma(t)}\\
  \frac{dp_{\psi}}{dt}&=&\frac{\dot{J}_2}{\gamma(t)}=
  \frac{\hat{\mathcal{M}}}{R_\text{NS}}\frac{c_{xy}}{\gamma(t)}\left(a_{2}^{2}-a_{1}^{2}\right)\,.\label{dppsi}
\end{eqnarray}
In the above equations, the $c_{ij}$'s are the PN tidal tensor components
defined later in Sec. \ref{sec:posttid},
\begin{equation}
\tilde A_i\equiv\int_0^\infty\frac{du}{(a_i^2+u)\sqrt{(a_1^2+u) (a_2^2+u) (a_3^2+u)}}
\label{defAi}
\end{equation}
are elliptic integrals,
\begin{eqnarray}
  \gamma(t)&=&1+\frac{\chi_{1}}{c^{2}}\left(
    \frac{Gm}{r}+\frac{v^{2}}{2}\chi_{1}\right)+\frac{\chi_{1}}{2c^{4}}
\bigg\{\frac{G^{2}m^{2}}{r^{2}}[1-4\chi_{2}]\nonumber\\
  &&+\frac{Gm}{r}v^{2}\left[5-\chi_{2}(1+2\chi_{1}\chi_{2}\right)]\nonumber\\
  &&+\chi_{1}\left[1-3\chi_{1}+\frac{11}{4}\chi_{1}^{4}\right]v^{4}\bigg\}\label{lorenztg}\\
  \hat{\mathcal{M}}&=&\frac{4\pi}{3}\int^{R_\text{NS}}_{0}\hat\rho\left(1-\frac{2V(r_s)}{c^2}
   +\frac{Gm(r_s)}{r_sc^2}\right) r^{4}_{s}dr_{s}\label{expm} \\
 \Pi&=&\frac{a_1a_2a_3}{R_\text{NS}^3}4\pi\!\!\int_0^{R_\text{NS}}\!\!\!\!\!\!\!\hat
 p\left(\hat\rho\frac{R_\text{NS}^3}{a_1a_2a_3}\right)
 \left(1\!+\!\frac{Gm(r_s)}{r_sc^2}\right)r_s^2dr_s\nonumber\\
 &&\label{exppi}
\end{eqnarray}
where $\chi_A=m_A/m$ ($A=1,2$), $V(r_s)\equiv
G\int^{\infty}_{r_s}\frac{m_s(r_s')}{r_s^{'2}}d{r'_s}$, $m_s(r_s)$ is the
gravitational mass enclosed in a sphere of radius $r_s$, $r$ and $v$ are the
orbital distance and the relative velocity of the two bodies in the PN frame,
and
\begin{eqnarray}
  \mathcal{C}&=&\frac{\hat{\mathcal{M}}}{R_\text{NS}^{2}}\left[\left(a_{1}^{2}
      +a_{2}^{2}\right)\Lambda-2a_{1}a_{2}\Omega\right]\,,\nonumber\\ 
  J_2&=&\frac{\hat{\mathcal{M}}}{R_\text{NS}^{2}}\left[\left(a_{1}^{2}
      +a_{2}^{2}\right)\Omega-2a_{1}a_{2}\Lambda\right]\label{CJLO}
\end{eqnarray}
are the circulation and the NS spin angular momentum. In this article we
consider the case of an asymptotically non-rotating star. As discussed in
\cite{WL00}, the intrinsic spin of the star may be defined only far away from
the companion, where the star is axisymmetric and the ellipsoid rotation
$\Omega$ vanishes; in this region, it is possible to identify the angular
velocity of the star with $-\Lambda$. Therefore, an asymptotically non-rotating
star has $\Lambda=0$, $\Omega=0$, and thus $\mathcal{C}=0$, at
$r\rightarrow\infty$. Since $\mathcal{C}$ is a constant of motion, it remains
zero during the inspiral. $\Omega$ and $\Lambda$, instead, become non-vanishing
as the NS couples with the tidal field, i.e., the ellipsoidal star acquires a
(very small) angular velocity associated with general relativistic effects, such
as geodetic precession and frame dragging.

\subsection{The orbital motion}\label{orbital}
The orbital motion is described in the post-Newtonian framework starting from
the the $3$PN metric of the two-body system, written in harmonic coordinates
$\{x^0=ct,x,y,z\}$ \cite{BFP98,FBA06}.  We assume an adiabatic inspiral of
quasi-circular orbits, i.e. such that the energy carried out by GWs is balanced
by the change of the total binding energy of the system \cite{BAF11}.  We use
the Taylor T4 approximant to describe the orbital phase evolution of the
two-body system, including the effects of the tidal interaction on the orbital
motion, derived at the $1$PN order (beyond the leading term) in
\cite{VFH11}. The phase $\phi(t)$ and the orbital frequency $\omega=d\phi(t)/dt$
are found by numerically integrating the following ordinary differential
equations:
\begin{eqnarray}
\frac{dx}{dt}&=&\mathcal{F}_\text{pp}+\mathcal{F}_\text{tid}\label{def:TaylorT4a}\\
\frac{d\phi}{dt}&=&\frac{c^{3}}{Gm}x^{3/2}\label{def:TaylorT4b}
\end{eqnarray}
where 
\begin{equation}\label{def:PNx}
x=\left(\frac{Gm\omega}{c^{3}}\right)^{2/3}\ ,
\end{equation}
$\mathcal{F}_\text{pp}$ is the point-particle term \cite{SOA10}, and
$\mathcal{F}_\text{tid}$ incorporates the finite-size effects on the orbital motion
\cite{VFH11}. To determine the radial coordinate $r(t)$, we employ the PN
expression for $\gamma=\frac{Gm}{rc^{2}}$ (not to be confused with the time
dilation factor $\gamma(t)$ defined in Eq.~(\ref{lorenztg})), which is also
known up to $3$PN order \cite{Fav11}. We refer to \cite{FGM11} for the explicit
form of the equations, or to the original papers \cite{BAF11,Fav11}.

\subsection{The post-Newtonian tidal tensor}\label{sec:posttid}
The quadrupolar tidal tensor is
\begin{equation} 
C_{ij}=R_{\alpha\beta\gamma\delta}e^{\alpha}_{(0)}e^{\beta}_{(i)}e^{\gamma}_{(0)}e^{\delta}_{(j)}\ ,
\end{equation} 
where $R_{\alpha\beta\gamma\delta}$ is the Riemann tensor of the $3$PN metric
describing the orbital motion. We introduce an orthonormal tetrad field
$e_{(i)}$ $(i=0,\dots,3)$ associated with the frame fixed to the star center of
mass $\mathcal{O}^*$, parallel transported along its motion, and such that
$e^{\mu}_{(0)}=u^{\mu}$, i.e., the $4-$velocity of $\mathcal{O}^*$. In
\cite{FGM11} we computed $C_{ij}$ up to $\mathcal{O}(1/c^3)$, whereas previous
computations were performed up to $\mathcal{O}(1/c^2)$ \cite{M83,DSX92}. In this
work we further expand the PN tidal tensor in order to include also $1/c^{4}$
terms. To this aim we first need to calculate the $2$PN component of the spatial
tetrad vectors $e^{j}_{(k)}$ introduced in \cite{F88}. From the orthogonality
condition $g_{\mu\nu}e^{\mu}_{(i)}e^{\nu}_{(j)}=\delta_{ij}$ we find that the
tetrad field expanded up to $1/c^{4}$ order for the bodies $A=1,2$ is:
\begin{eqnarray}
  e^{t}_{(t)A}&=&{\tilde e}^{t}_{(t)A}\nonumber \\
  e^{j}_{(t)A}&=&{\tilde e}^{j}_{(t)A}\nonumber\\
  e^{t}_{(j)A}&=&{\tilde e}^{t}_{(j)A}\nonumber\\ 
  e^{j}_{(x)A}&=&{\tilde e}^j_{(x)A}\cos\xi_{A}+{\tilde e}^j_{(y)A}\sin\xi_{A}\nonumber\\ 
  e^{j}_{(y)A}&=&-{\tilde e}^j_{(x)A}\sin\xi_{A}+{\tilde e}^j_{(y)A}\cos\xi_{A}\nonumber\\ 
  e^{j}_{(z)A}&=&{\tilde e}^j_{(z)A}\nonumber
\end{eqnarray} 
with
\begin{widetext}
  \begin{eqnarray}\label{tet} {\tilde
      e}^{t}_{(t)A}&=&1+\frac{1}{c^{2}}\left[(V)_{A}+\frac{v_{A}^{2}}{2}\right]+\frac{1}{c^{4}}\left[\frac{5}{2}
      v_{A}^{2}(V)_{A}+\frac{1}{2}\left(V^{2}\right)_{A}
      +\frac{3}{8}v_{A}^{4}-4v_{A}^{i}(V_{i})_{A}\right]
    +\mathcal{O}(c^{-6})\\
    \nonumber 
    {\tilde e}^{j}_{(t)A}&=&\frac{v_{A}^{j}}{c}+\left[(V)_{A}+\frac{v_{A}^{2}}{2}\right]
    \frac{v_{A}^{j}}{c^{3}}+\mathcal{O}(c^{-5})\\
    \nonumber 
    {\tilde e}^{t}_{(j)A}&=&\frac{v_{A}^{j}}{c}+\frac{1}{c^{3}}\left[-4(V_{j})_{A}+v_{A}^{j}\left(3(V)_{A}
        +\frac{v_{A}^{2}}{2}\right)\right]+\mathcal{O}(c^{-5})
    \nonumber\\ 
    {\tilde e}^{j}_{(k)A}&=&\delta^{j}_{k}\left[1-\frac{(V)_{A}}{c^{2}}\right]+\frac{v_{A}^{j}v_{A}^{k}}{2c^{2}}
    +\frac{1}{8c^{4}}\left[4(V)_{A}^{2}\delta^{j}_{k}+3v_{A}^{j}v_{A}^{k}\left(4(V)_{A}+v_{A}^{2}\right)
      -16(\hat{W}_{jk})_{A}\right]+\mathcal{O}(c^{-6})\ ,
    \nonumber
  \end{eqnarray} 
\end{widetext}
where $(V)_{A}$, $(V_{i})_{A}$, $(\hat{W}_{ij})_{A}$ are PN potentials
defined in terms of the source densities, evaluated at the location of
the body $A$, and regularized as described in \cite{B06}; finally,
$\xi_{A}$ is the angle describing geodesic precession and frame
dragging
\begin{equation}
  \xi_{A}=\frac{\left({\cal Q}_{xy}\right)_{A}}{c^{2}}\left[1+\frac{1}{c^{2}}\left(\left(V\right)_{A}-\frac{v_{A}^{2}}{4}\right)\right]\,;
\end{equation} 
here ${\cal Q}_{xy}$ is the non vanishing component of the antisymmetric matrix
${\cal Q}$ defined as \cite{F88}:
\begin{equation}
  {\cal Q}(t,t_{0})=\int^{t}_{t_{0}}\left[\textbf{v}\times{(\nabla V-\textbf{a})}
    -\nabla\times(V\textbf{v}-2{\textbf{V}})\right]dt\,,
\end{equation} 
with $t_{0}$ arbitrary initial time.

Following the same procedure described in \cite{FGM11} we project the Riemann
tensor on the tetrad field, to derive the tidal tensor which we express in the
NS principal frame as $c=TCT^{T}$, where
\begin{equation}
  T=\left(\begin{array}{ccc}
      \cos\psi & \sin\psi &0\\
      -\sin\psi & \cos\psi & 0\\
      0 & 0 &1    
\end{array}\right)
\end{equation}
and $\psi$ has been defined in Section \ref{sec:aff}. We find that the tidal
tensor components for the body $2$, appearing on the right-hand-side of the
dynamical equations (\ref{eqa1})-(\ref{eqa3}), are:
\begin{widetext}
  \small{
    \begin{eqnarray}
      c_{xx}=&-&\frac{Gm\chi_1}{2r^3}\left\{1+3\cos[2\psi_{l}]\right\}+\frac{Gm\chi_{1}}{4c^2 r^3}
      \left\{\frac{Gm}{r}(5+\chi_{1})\left(1+3\cos[2\psi_{l}]\right)-6v^{2}\left(1+\cos[2\psi_{l}]\right)\right\}
      +\frac{3G }{c^3}\frac{S_{1}^{z}\dot{\phi}}{r^{3}}\left\{\delta \chi\ +\right.\nonumber\\
      &+&\left.(\chi_{1}-5\chi_{2})\cos[2\psi_{l}]\right\}+\frac{G m\chi_{1}}{c^{4}r^3} 
      \Bigg\{\bigg[\frac{G^2 m^2}{r^{2}} \left(\frac{687}{28}\chi_{2}^2-\frac{101}{8}\chi_{2}-9\right)-\frac{G m}{r}v^{2} 
      \left(3\chi_{2}^{2}+\frac{223}{16}\chi_{2}+\frac{3}{2}\right)\ +\nonumber\\
      &-&\frac{3}{2} v^{4}\left(2\chi_{2}^2-2\chi_{2}+1\right)\bigg]
      \cos [2\psi_{l}]+\frac{G^2 m^2}{r^2}
      \left(\frac{229}{28}\chi_{2}^{2}+\frac{\chi_{2}}{8}-3\right)-
      \frac{G m}{r}v^{2}\left(\chi_{2}^{2}+\frac{109}{16}\chi_{2}+\frac{3}{2}\right)\ +\nonumber\\
      &-&\frac{3}{2}
      \left(2\chi_{2}^2-2\chi_{2}+1\right)v^{4}\bigg\}+\mathcal{O}(c^{-5})\label{cxx}\ ,\\
      \phantom{a}\nonumber\\
      c_{yy}=&-&\frac{Gm\chi_{1}}{2r^3}\left\{1-3\cos[2\psi_{l}]\right\}+\frac{Gm\chi_{1}}{4 c^2
        r^4}\left\{\frac{Gm}{r}\left(5+\chi_{1}\right)(1-3\cos[2\psi_{l}])-6v^{2}(1-\cos[2\psi_{l}])\right\}+
      \frac{3G}{c^{3}}\frac{S_{1}^{z}\dot{\phi}}{r^3}\big\{\delta \chi\ +\nonumber\\
      &-&(\chi_{1}-5\chi_{2})\cos[2\psi_{l}]\big\}+\frac{G m\chi_{1}}{c^{4}r^3} 
      \Bigg\{\bigg[-\frac{G^2 m^2}{r^{2}} \left(\frac{687}{28}\chi_{2}^2-\frac{101}{8}\chi_{2}-9\right)+\frac{G
        m}{r}v^{2}\left(3\chi_{2}^{2}+\frac{223}{16}\chi_{2}+\frac{3}{2}\right)\ +\nonumber\\
      &+&\frac{3}{2} \left(2\chi_{2}^2+2\chi_{2}+1\right)v^{4}\bigg]\cos [2
      \psi_{l}]+\frac{G^2 m^2}{r^2} 
      \left(\frac{229}{28}\chi_{2}^{2}+\frac{\chi_{2}}{8}-3\right)-\frac{G
        m}{r}v^{2}\left(\chi_{2}^{2}+
        \frac{109}{16}\chi_{2}+\frac{3}{2}\right)\ +\nonumber\\
      &-&\frac{3}{2}
      \left(2\chi_{2}^2-2\chi_{2}+1\right)v^{4}\bigg\}+\mathcal{O}(c^{-5})\label{cyy}\ ,\\
      \phantom{a}\nonumber\\
      c_{zz}=&&\frac{Gm\chi_{1}}{r^3}-\frac{Gm\chi_{1}}{2c^2
        r^{3}}\left[\frac{Gm}{r}\left(5+\chi_{1}\right)-
        6v^{2}\right]-\frac{6G}{c^3}\frac{\dot{\phi}S_{1}^{z}}{r^3}\delta\chi
      +\frac{G m\chi_{1}}{c^4r^3} \Bigg\{\frac{G^2 m^2}{r^{2}} 
      \left[-\frac{229}{14}\chi_{2}^{2}+\frac{51}{4}\chi_{2}+6\right]\ +\nonumber\\
      &+&\frac{G m}{r}v^{2}\left[2\chi_{2}^{2}+\frac{5}{8}\chi_{2}+3\right]+
      3\left[2\chi_{2}^2-2\chi_{2}+1\right]v^{4}\Bigg\}+\mathcal{O}(c^{-5})\label{czz}\ ,\\
      \phantom{a}\nonumber\\
      c_{xy}=&&\frac{3Gm\chi_{1}}{2r^{3}}\sin[2\psi_{l}]-\frac{3Gm\chi_{1}}{4 c^2 r^3}
      \left[\frac{G  m}{r}\left(5+\chi_{1}\right)
        -2 v^{2}\right] \sin[2\psi_{l}]+\frac{3G}{c^{3}}\frac{ 
        S_{1}^{z}\dot{\phi}}{r^3}(5\chi_{2}-\chi_{1})\sin[2\psi_{l}] +\frac{G m\chi_{1}}{ 
        c^{4}r^3}\times\nonumber\\
      &\times& \bigg\{\frac{G^{2}m^2}{r^{2}}
      \left[-\frac{687}{28}\chi_{2}^{2}+\frac{101}{8}
        \chi_{2}+9\right]+\frac{Gm}{r}v^{2}
      \left[3\chi_{2}^{2}+\frac{223}{16}\chi_{2}+\frac{3}{2}\right]+\frac{3}{2} 
      \left[2\chi_{2}^2-2\chi_{2}+1\right]v^{4}\Bigg\}\sin [2\psi_{l}]+\mathcal{O}(c^{-5})\label{cxy}\ .
    \end{eqnarray}}
\end{widetext}
In the above expressions, $\delta \chi=\chi_{1}-\chi_{2}$ and the \emph{lag}
angle $\psi_{l}=\psi-\phi+\xi$ describes the misalignment between the $a_{1}$
axis and the line between the two objects. The superscript dot identifies
derivatives of the orbital variables $\phi$ with respect to the coordinate time
$t$. It should be noted that in the center-of-mass frame of the binary system,
the tidal tensor is given by the same expressions (\ref{cxx})-(\ref{cxy}),
setting $\psi_l=-\phi$.  We also stress that, by construction, the tidal tensor
is traceless.

Finally, we remark that (for a star with zero circulation, i.e.,
$\mathcal{C}=0$) the lag angle is very small, since it is only due to general
relativistic effects such as geodetic precession and frame dragging; therefore,
$\psi_l\ll1$. Then, since at leading order $c_{xy}\propto\sin[2\psi_l]$, it
follows that
\begin{equation}
c_{xy}\ll c_{xx},c_{yy},c_{zz}\,.\label{noxy}
\end{equation}
We have checked numerically that $c_{xy}$ is always several orders of magnitude
smaller than the others components of the tidal tensor. We remark that, as noted
in \cite{LRS94}, the lag angle becomes non-negligible when viscosity is included
in the model.

\section{Tidal Love numbers in the post-Newtonian-affine
  approach}\label{sec:love}
NS tidal deformations can be described in terms of a set of parameters, the Love
numbers, encoding the deformation properties of the star. The idea is that in
presence of a weak, external tidal field, a spherical star is deformed, and its
mass multipole moments are proportional to the multipole moments of the
perturbing tidal field. In this paper, we focus on the Love number $k_2$, which
is associated with the lowest order ``electric'' moment, i.e., the $\ell=2$, or
quadrupole, moment. This is the most relevant for the phenomenology of stellar
deformations, to be considered when modelling gravitational waveforms. The mass
quadrupole moment (traceless) tensor is proportional to the tidal field
\cite{FH08,H08}:
\begin{equation}
\label{love}
Q_{ij}=-\frac{2}{3}k_2R_\text{NS}^5C_{ij}\,.
\end{equation}  
In the NS principal frame, this equation reads
\begin{equation}  
  q_{ij}=-\frac{2}{3}k_2R_\text{NS}^5c_{ij}\label{ad1}\,,
\end{equation}  
where $q_{ij}$ is the quadrupole moment tensor projected onto this frame
\begin{eqnarray}
  q_{ij}&=&\int
  dM_B\left(x^ix^j-\frac{1}{3}\delta^{ij}r^2\right)d^3x\nonumber\\
  &=&\frac{\hat{\cal M}}{R_\text{NS}^2}(a_i^2-a^2)\delta^{ij}
\label{ad2}
\end{eqnarray}
(no sum on $i$), $a^2=(a_1^2+a_2^2+a_3^2)/3$, and $\hat{\mathcal M}$ is the
scalar quadrupole moment for the spherical, isolated star configuration given in
Eq.~(\ref{expm}).

Eq.~(\ref{ad1}) is based on the adiabatic approximation, which assumes that the
timescale of the orbital evolution (and thus of tidal tensor variations) is much
larger then the timescale needed for the star to readjust into a stationary
configuration. In the following we will drop this assumption and assess its
validity.

\subsection{Evolution of the tidal Love number during the inspiral}

We solve Eqns.~(\ref{da1})-(\ref{dppsi}), (\ref{def:TaylorT4a}), and
(\ref{def:TaylorT4b}), for a representative set of binary system
configurations. We consider two equations of state, dubbed APR (which had been
derived by Akmal, Pandharipande and Ravenhall [37]) and PS (which had been
derived by Pandharipande and Smith [38]). These EOSs are expected to cover the
range of possible EOS stiffness and NS deformability:
the APR EOS describes soft NS matter and yields NS models with
high compactness ${\cal C}$ (not to be confused with the circulation
defined in Eq.~\ref{CJLO}) and small deformability $\lambda$, whereas the PS
EOS describes stiff NS matter and gives stellar models with small compactness and
large deformability (see for instance \cite{PROR11}). We consider two values of
the secondary NS mass, $m_2=1.2\,M_\odot,1.4\,M_\odot$, and three values of the
mass ratio, $q=m_1/m_2=1,3,5$. In the $q=1$ cases, both bodies are NSs, with the
same EOS, and we take into account the deformation of both of them. When $q=3,5$
we assume that primary body is a BH.

For each of these binary models, we compute $k_2$ from Eq.~(\ref{ad1}), using
Eq.~(\ref{ad2}) for the left-hand side. We find that this quantity is a function
of the orbital distance, $k_2(r)$, and that it increases during the stellar
inspiral. We express $k_2$ as a function of $r$, and not of the PN variable $x$,
because  it is easier to find an accurate fit for $k_2$ in terms of
the $r$-dependence rather than of the $x$-dependence. Further, we think that the
variable $r$ provides a more direct physical insight on the process. Hereafter,
we denote the asymptotic limit of the function $k_2(r)$, i.e., the Love number,
as
\[
\bar k_2=\lim_{r\rightarrow\infty} k_2(r)\,.
\]
We determine $\bar k_2$ by fitting our data with the following expression:
\begin{equation}\label{lovefit}
k_{2}(r)=\bar{k}_{2}\left(1+\alpha\frac{m}{r}+\beta\frac{m^{2}}{r^{2}}
+\gamma\frac{m^{3}}{r^{3}}\right)\,.
\end{equation}
\begin{figure*}[t]
\epsfig{file=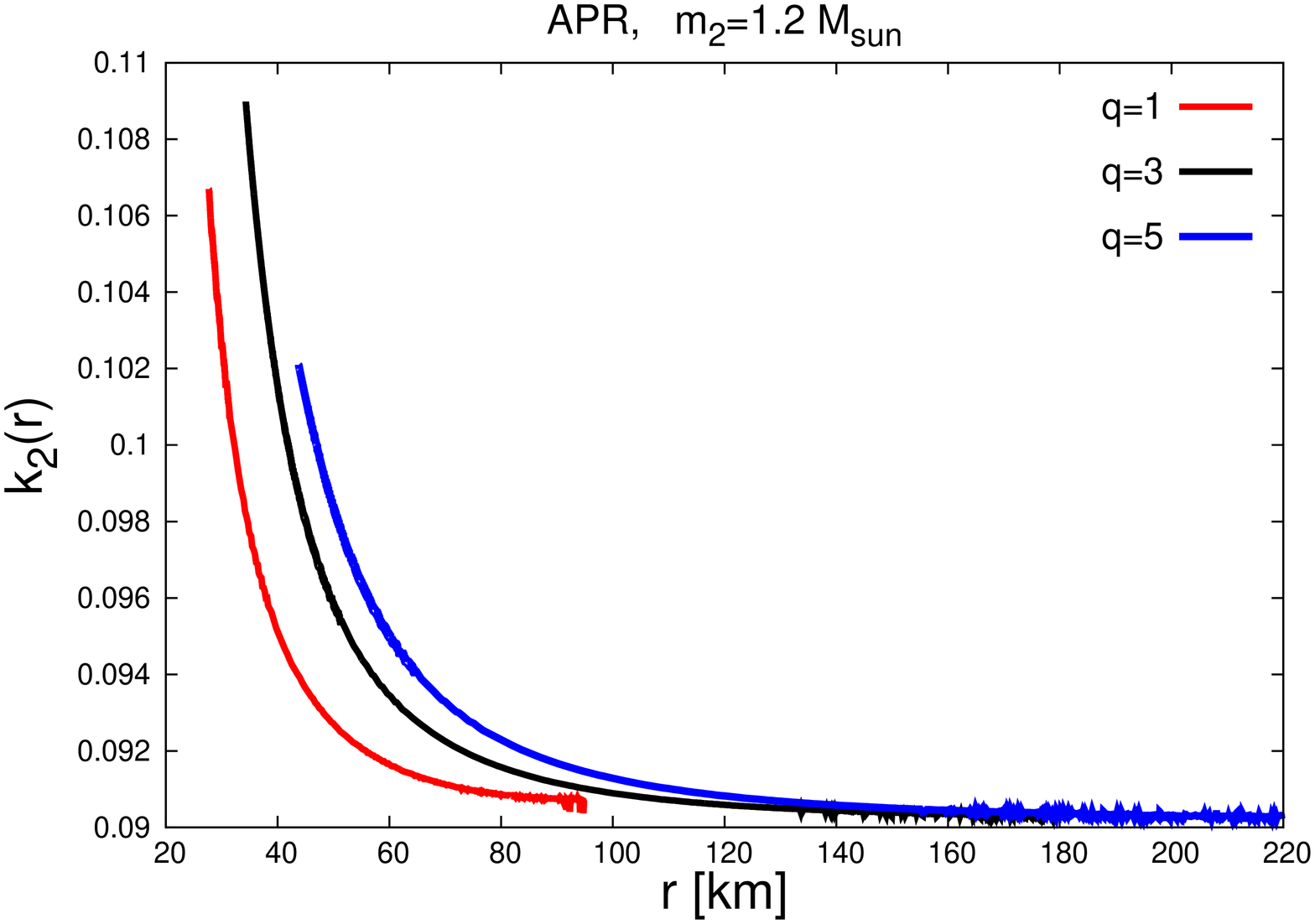,angle=-90,width=240pt}
\epsfig{file=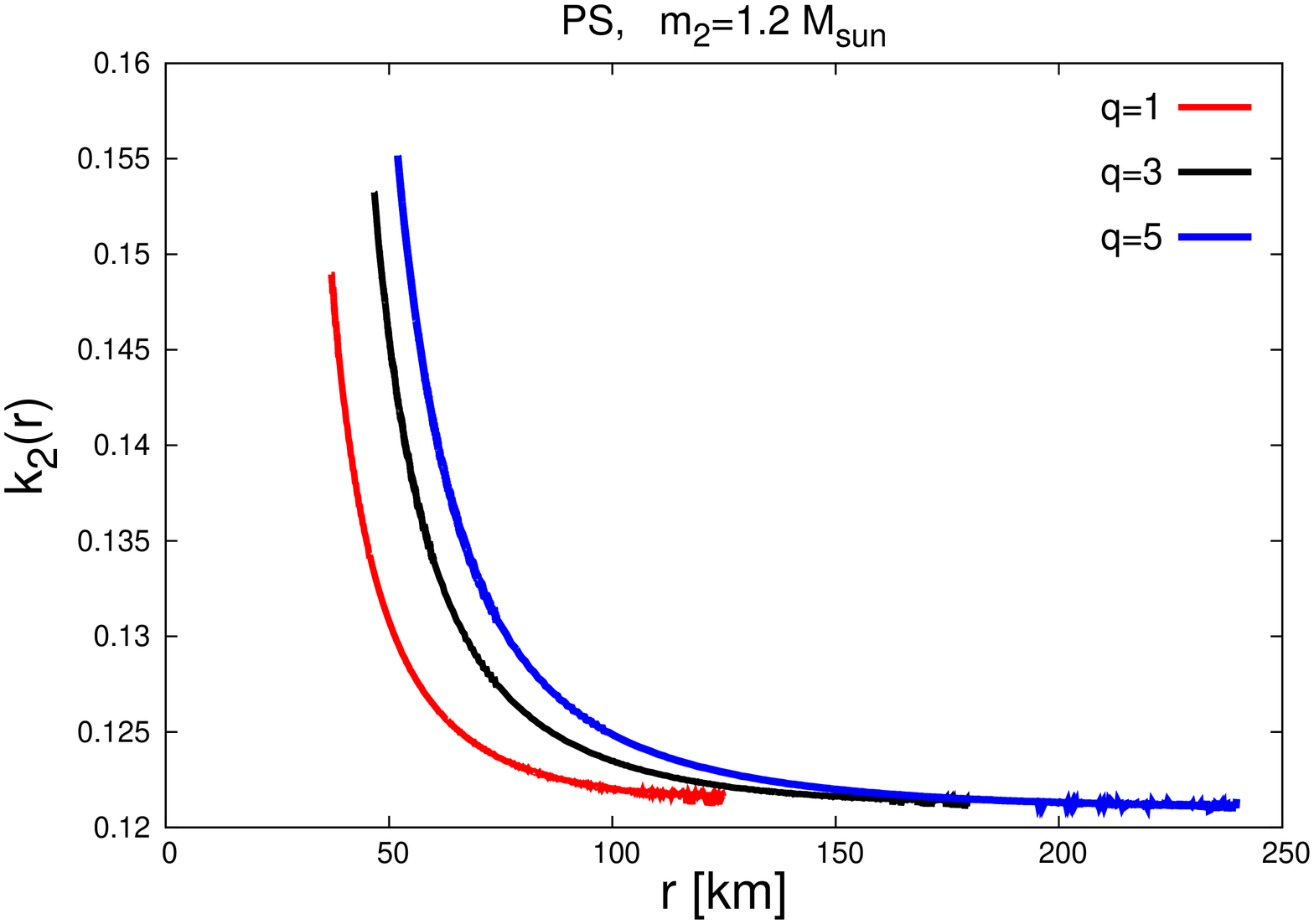,angle=-90,width=240pt}\\
\vspace{0.5cm}
\epsfig{file=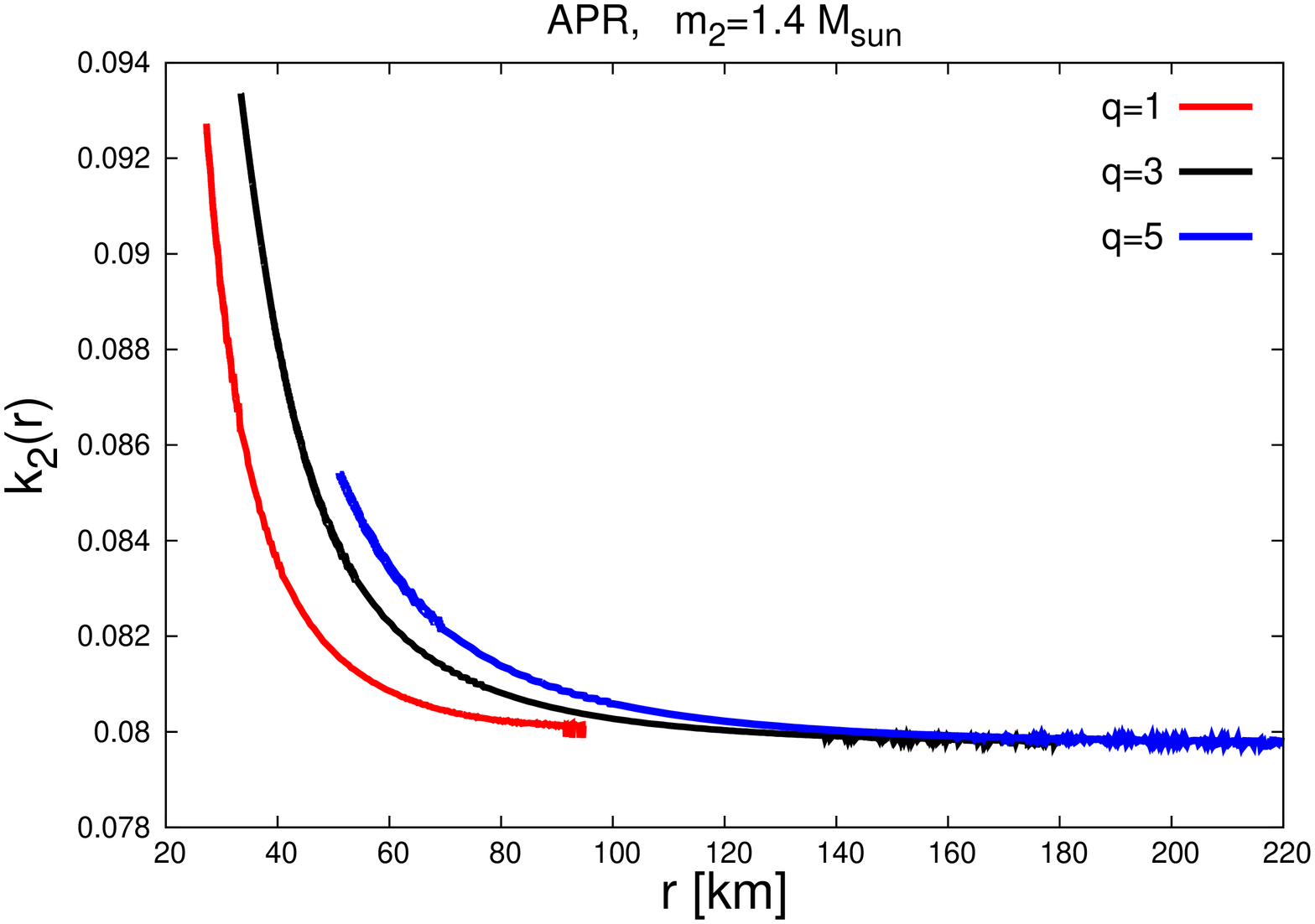,angle=-90,width=240pt}
\epsfig{file=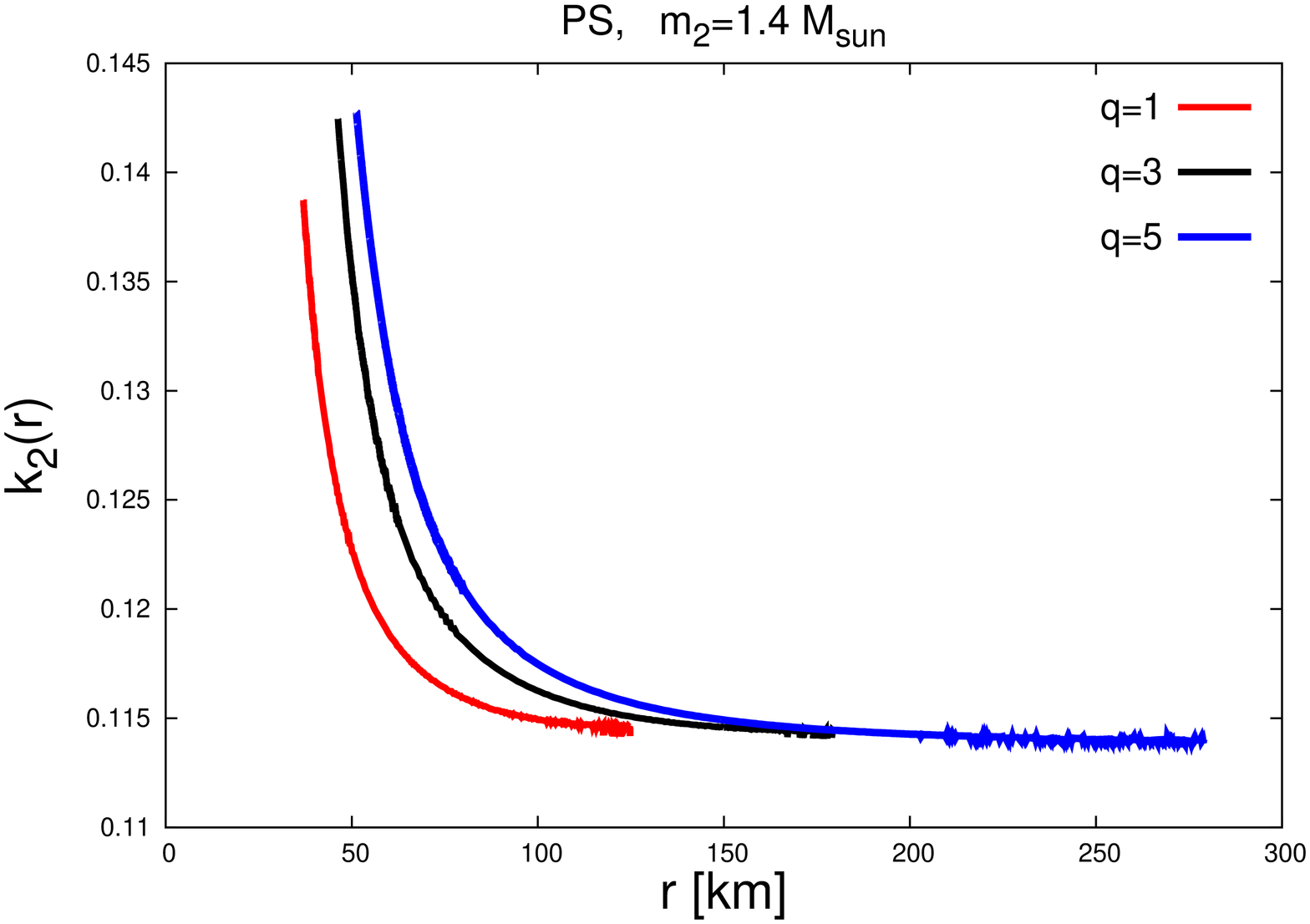,angle=-90,width=240pt}\\
\caption{(Color online) \label{k2_r} Love function $k_2(r)$ for
  different choices of the NS EOS, NS mass $m_2$, and binary mass
  ratio $q$, as indicated in the panels. For the APR EOS, $q=5$, and
  both values of the NS mass, the NS reaches the innermost circular
  orbit before the mass-shedding sets in.}
 \end{figure*}

 In Fig.~\ref{k2_r} we plot the function $k_2(r)$ for all the binary
 configurations we considered. The figure shows that in the last stages of the
 inspiral, before the mass-shedding sets in, $k_2(r)$ grows by a factor of
 $\sim10\%-30\%$. This effect is strongly dependent on the EOS choice: it is
 larger/smaller for a stiffer/softer EOS as PS/APR, or, equivalently, for a
 less/more compact NS. The dependence of this effect on the mass ratio $q$ is
 weaker; the Love number increases more for larger values of the mass ratio,
 unless the star reaches the innermost circular orbit (ICO) before the
 mass-shedding sets in. It seems that this effect is nearly insensitive to the
 NS mass. We show the curves in Fig.~\ref{k2_r} as functions of $r$, and not of
 $r/m$, because we find that this choice makes the common asymptotic limit of
 the curves more evident.

\subsection{Equations of tidal deformation in the perturbative
  regime}\label{sec:etd}
In the limit of small deformations
\begin{equation}
  \delta a_i^2\equiv a_i^2-R_\text{NS}^2\ll a^2\,,\label{smalldef}
\end{equation}
it is possible to derive a semi-analytical, closed form for the Love number
$\bar k_2$, expressed in terms of the NS scalar quadrupole moment and of the
integral of the pressure over the stellar volume.

By introducing the notation $\delta a^2\equiv\frac{1}{3}\sum_i\delta
a_i^2=a^2-R_\text{NS}^2$, we rewrite Eq.~(\ref{ad2}) as
\begin{equation}
  q_{ij}=\frac{\hat{\cal M}}{R_\text{NS}^2}(\delta a_i^2-\delta a^2)\delta^{ij}\,\label{pertq2}
\end{equation}
(no sum on $i$). As a first step, we show that the spherical, non rotating
solution $a_i=R_\text{NS}$, $\Omega=\Lambda=0$, is a solution of
Eqns.~(\ref{da1})-(\ref{dppsi}), with $c_{ij}=0$. Indeed, in this limit the
dynamical equations reduce to
\begin{equation}
  \frac{\hat{\mathcal{M}}}{\gamma(t)}
  \left[\frac{1}{2}\frac{\hat{\mathcal{V}}}{\hat{\mathcal{M}}}R_\text{NS}^3\tilde A+
    \frac{\hat\Pi}{\hat{\mathcal{M}}}\right]=0~,\label{eqsph}
\end{equation}
where we introduced $\tilde A\equiv\tilde A_i(a_i=R_\text{NS})$, which from
Eq.~(\ref{defAi}) thus reads
\begin{equation}
  \tilde A=\int_0^\infty\frac{du}{(R_\text{NS}^2+u)^{5/2}}=\frac{2}{3}\frac{1}{R_\text{NS}^3}\,.
\end{equation}
Eq.~(\ref{eqsph}) is therefore satisfied when Eq.\,(\ref{virial}),
$\hat{\mathcal{V}}=-3\hat\Pi$, i.e., the virial theorem, holds.

We now consider the first order perturbative expansion of
Eqns.~(\ref{da1})-(\ref{dppsi}) around the spherical, non-rotating
configuration. This expansion is an accurate description of the dynamical system
when deformations are small, i.e., when the tidal field is weak and the rotation
rate is small.
 
To simplify the discussion, we restrict the discussion to the case of an
asymptotically non-rotating star; in this case Eq.~(\ref{noxy}) holds, i.e.,
$c_{xy}\ll c_{xx},c_{yy},c_{zz}$, and Eqns.~(\ref{dlambda})-(\ref{dppsi}) and
(\ref{CJLO}) guarantee that $\Omega$, $\Lambda$, $\lambda$, and $\psi$ may be
neglected. The remaining non-trivial equations, (\ref{eqa1})-(\ref{eqa3}),
reduce to
\begin{equation}
 \frac{1}{2}\frac{\hat{\cal V}}{\hat{\mathcal{M}}}R_\text{NS}^{3}\tilde{A}_{i}+
  \frac{R_\text{NS}^{2}}{\hat{\mathcal{M}}}\frac{\Pi}{a_{i}^{2}}-c_{ii}
=0\label{pert1}
\end{equation}
($i=1,2,3$, no sum on $i$). We remark that in this proof the explicit expression
of the tidal field $c_{ij}$, which drives the stellar deformation, can be taken
as generic (we only require that $c_{xy}\ll c_{xx},c_{yy},c_{zz}$).

Expanding Eq.\,(\ref{pert1}) we find
\begin{equation}
\frac{\hat{\mathcal{V}}}{2\hat{\mathcal{M}}}\left(\frac{2}{3}-\frac{2}{5}\frac{\delta
    a_i^2}{R_\text{NS}^2}\right)+\frac{\hat\Pi}{\hat{\mathcal{M}}}\left(1+\frac{\delta\Pi}{\hat\Pi}
-\frac{\delta a_i^2}{R_\text{NS}^2}\right)-c_{ii}=0\,,
\end{equation}
where we denoted the expansion of $\Pi$ as $\hat\Pi+\delta\Pi$ and considered
the Taylor expansion of the integrals $\tilde A_i$ (\ref{defAi}) at first order
in $\delta a_i^2=a_i^2-R_\text{NS}^2$, i.e.,
\begin{equation}
  \tilde A_i=\frac{2}{3R_\text{NS}^3}-\frac{2}{5}\frac{\delta a_i^2}{R_\text{NS}^5}\,.
\end{equation}
Imposing the virial theorem $\mathcal{V}=-3\hat\Pi$, one now has
\begin{equation}
\frac{\hat\Pi}{\hat{\mathcal{M}}}\left(\frac{\delta\Pi}{\hat\Pi}
-\frac{2}{5}\frac{\delta a_i^2}{R_\text{NS}^2}\right)-c_{ii}=0\,.\label{pert2}
\end{equation} 
The traceless part of Eq.~(\ref{pert2}) yields
\begin{equation}
  \frac{2}{5}\frac{\hat\Pi}{\hat{\mathcal{M}}}\frac{\delta a_i^2-\delta a^2}{R_\text{NS}^2}=-c_{ii}\label{pert3}
\end{equation}
and plugging this into Eq.~(\ref{pertq2}) gives
\begin{equation}
  q_{ij}=\delta_{ij}\frac{\hat{\mathcal{M}}}{R_\text{NS}^2}(\delta
  a^2_i-\delta
  a^2)=-\frac{5}{2}\frac{\hat{\mathcal{M}}^2}{\hat\Pi}c_{ij}\,.
\label{LC}
\end{equation}
A similar relation was found in \cite{CL83}.

Finally, Eqns.~(\ref{ad1}) and (\ref{LC}) yield the semi-analytical expression 
for the Love number we were looking for:
\begin{equation}
  \bar k_2=\frac{15}{4}\frac{\hat{\mathcal{M}}^2}{\hat\Pi R_\text{NS}^5}\,,\label{k2an}
\end{equation}
where $\hat\Pi$ and $\hat{\mathcal{M}}$ are given in Eqns.~(\ref{exppi}) and
(\ref{expm}), respectively.
\begin{table}[ht]
  \centering
  \begin{tabular}{ccccccccccc}
    \toprule
    EOS & $m_{2}(M_{\odot})$ & $q$ && $\bar{k}_{2}$ && $\Delta\bar{k}_2/\bar k_2$ &&
 $\Delta\bar{k}^{an}_2/\bar k_2$ &&$f^{orb}_{cut}(Hz)$ \\  
    \colrule
    APR & $1.2$ & $1$ & \phantom{a} & $0.0884$ && $0.011$ &&$0.031$&& $532.43$\\
    & $1.2$ & $3$ && $0.089$ && $0.019$ && && $500.32$\\
    & $1.2$ & $5$ && $0.090$ && $0.033$ && &&$404.88^\dagger$\\
    & $1.4$ & $1$ && $0.079$ && $0.047$ &&$0.058$&&$575.88$\\
    & $1.4$ & $3$ && $0.081$ && $0.071$ && &&$541.07$\\
    & $1.4$ & $5$ && $0.080$ && $0.063$ && &&$347.07^\dagger$\\
    \colrule
    PS & $1.2$ & $1$ && $0.117$ && $0.024$ &&$0.006$&& $354.81$\\
    & $1.2$ & $3$ && $0.114$ && $0.048$ && &&$332.39$\\
    & $1.2$ & $5$ && $0.116$ && $0.034$ && &&$325.55$\\
    & $1.4$ & $1$ && $0.110$ && $0.017$ &&$0.017$&&$378.04$\\
    & $1.4$ & $3$ && $0.108$ && $0.032$ && &&$354.11$\\
    & $1.4$ & $5$ && $0.112$ && $0.005$ && &&$345.20$\\
    \botrule
  \end{tabular} 
  \caption{Values of the Love number for different EOSs, NS masses $m_{2}$,
    and binary mass ratios $q$ (columns 1,2,3, respectively). The value of
    $\bar{k}_{2}$ given in column 4 is computed solving
    Eqns.~(\ref{da1})-(\ref{dppsi}), (\ref{def:TaylorT4a}), and
    (\ref{def:TaylorT4b}); in columns 5 and 6 we show the relative errors
    $(\bar{k}_{2}-{k}^\text{H}_{2})/\bar{k}_{2}$ and 
    $(\bar{k}_{2}^\text{an}-{k}^\text{H}_{2})/\bar{k}_{2}$, where
    $\bar{k}_{2}^\text{H}$ is the relativistic Love number computed
solving the relativistic equations of stellar perturbations
\cite{H08}, and $\bar{k}_{2}^\text{an}$ is evaluated
    using the semi-analytic formula~(\ref{k2an}) (see text). The values of
    the orbital frequency $f^\text{orb}_\text{cut}$ at which our simulations
    end are also provided in the last column. The $^{\dagger}$ symbol indicates
    that the ICO is reached before the onset of the mass-shedding.
    \label{tablelove}}
\end{table}

In Table \ref{tablelove} we show the quantities characterizing the different
binary models considered in this paper, and the corresponding values of the Love
number $\bar k_2$, computed in three different ways: (i) from the dynamical
evolution of our equations (\ref{da1})-(\ref{dppsi}), (\ref{def:TaylorT4a}),
(\ref{def:TaylorT4b}); (ii) by solving the equations of relativistic stellar
perturbations for an isolated NS, derived in \cite{H08} ($\bar k_2^\text{H}$);
(iii) from the semi-analytical formula (\ref{k2an}) ($\bar k_2^\text{an}$). We
notice that, as expected (see also \cite{FGM11}, where polytropic EOSs were
considered), the less relativistic the NS is, i.e. the lower its compactness is,
the lower the relative error between the relativistic value $k_2^\text{H}$ and
our perturbative result $\bar k_2^\text{an}$ is. Note that when the Love number
is extracted from the dynamical evolution of the binary system, it has a (weak)
dependence on the mass ratio $q$; the quantities $\bar k_2^H$, $\bar k_2^{an}$,
instead, do not depend on $q$, because they are evaluated in terms of the
intrinsic properties of the star. We find that the values of $\bar k_2$ computed
with these three approaches have very small discrepancies, of at most few
percentage points.

Table \ref{tablelove} also provides the value of the orbital frequency
$f^\text{orb}_\text{cut}$ where our simulations stop. This corresponds to the
onset of mass-shedding, or to the ICO, if the latter is encountered before
mass-shedding (see discussion in \cite{FGM11}).

\section{Gravitational waveform in the post-Newtonian-affine
  approach}\label{sec:data}
We now compute the gravitational waveform including tidal effects by means of
the Love function $k_2(r)$. In the following derivation, we use geometric units
$G=c=1$. Our starting point is the state-of-the-art inclusion of tidal effects
in the GW signal by means of the Love number $\bar k_2$.
This is based on the binding energy ${\cal
  E}(x)$ and the gravitational flux ${\cal L}_{GW}(x)$ \cite{SOA10,VFH11}:
\begin{eqnarray}
  \mathcal{E}(x)=&-&\frac{m\nu x}{2}\left\{\sum_{k=0}^{6}e_{k}x^{k/2}-
    \bar{\lambda}_2\frac{\chi_1}{\chi_2}\frac{x^{5}}{m^{5}}\ \times\right.\nonumber\\
  &\times&\left.\left[9+\frac{11}{2}
      \left(3+2\chi_2+3\chi_2^{2}\right)x\right]+1\leftrightarrow 2\right\},\label{def:Energy}
\end{eqnarray}
\begin{eqnarray}
  \mathcal{L}_{GW}(x)=&&\frac{32}{5}\nu^{2}x^{5}\left\{\sum_{k=0}^{7}f_{k}x^{k/2}+
\frac{\bar{\lambda}_2}{\chi_2}\frac{x^5}{m^5}\times\right.\nonumber\\  
  &&\times\left. \left[6\left(3-2\chi_2
      \right)-\frac{1}{28}\left(704+1803\chi_2\ +\right.\right.\right.\nonumber\\
  &&-\left.\left.\left.4501\chi_2^{2}+2170\chi_2^{3}\right)x\right]
    +1\leftrightarrow 2\right\}\,,\label{def:Flux}
\end{eqnarray}
where $\bar{\lambda}_A$ ($A=1,2$) is the tidal deformability associated with the
$A$-th body and related to the Love number $\bar{k}_{2,A}$ by
\begin{equation}
\label{lambdaPP}
\bar{\lambda}_A=\frac{2}{3}\bar{k}_{2,A}R_\text{NS}^{5}\,.
\end{equation}
The point-particle coefficients $e_{k},f_{k}$ are given in the Appendix of
\cite{SOA10}.

Using the stationary phase approximation (SPA) \cite{SD91,CF94} and the TaylorF2
framework \cite{DIS01} to construct the signal in the frequency domain, the
$(l,m_l)$ mode of the GW signal reads
\begin{equation}
  \tilde{h}^{lm_{l}}\simeq\sqrt{\frac{2\pi}{m_{l}\ddot{\phi}(t_{f})}}A^{lm_{l}}e^{i\psi^{lm_{l}}(f)}\,,
\end{equation}
where $t_{f}$ is defined as the time when the instantaneous frequency matches
the Fourier variable, i.e. $m_{l}\omega(t_{f})=2\pi f$, $\phi$ is the orbital
phase, the dots indicate a second derivative with respect to time, $A^{lm_l}$ is
the Fourier amplitude, and $\psi^{lm_{l}}=2\pi
ft_f-m_{l}\phi(t_f)-\frac{\pi}{4}$ is the Fourier phase.  To compute the
$\ddot{\phi}=\dot{\omega}$ entering the GW signal, we follow the strategy
adopted in \cite{SOA10} and express $\dot\omega$ as
$\dot{\omega}=\frac{3}{2m}\sqrt{x}\dot{x}$, with $\dot{x}$ derived using the TaylorT4
prescription given by Eq.~(\ref{def:TaylorT4a}). Hereafter, we shall focus on
the $l=m_l=2$ mode of the gravitational waveform and drop the superscript ``22''
for sake of simplicity. The coefficients of the $3$PN order expansion of the
amplitude $A(x)$ are collected in \cite{SOA10}. The Fourier phase has the form
\begin{equation}
  \psi=\psi_\text{PP}+\psi_\text{T}\,,\label{psi_freq}
\end{equation}
where the coefficients of the $3.5$PN order expansion of the point-particle
contribution $\psi_\text{PP}(x)$ are given in \cite{SOA10} and the tidal term,
$\psi_\text{T}$, is calculated up to $1$PN (relative) order in
\cite{FH08,HLLR10,VFH11}, by assuming a constant Love number. It reads
\begin{equation}
  \psi_\text{T}= \psi^\text{N}_\text{T}+\psi^{1\text{PN}}_\text{T}\,,
\end{equation}
with
\begin{eqnarray}
\label{psi_T0}
\psi^\text{N}_\text{T}(f)&=& -\frac{3x^{5/2}}{128\nu m^{5}}\bar\lambda_2
\left[
  \frac{24}{\chi_2}\left(1+11\chi_1\right)\right]+
1\leftrightarrow 2\,,
\nonumber\\
&&\\
\psi^{1\text{PN}}_\text{T}(f)&=&-\frac{3x^{5/2}}{128\nu m^{5}}\bar\lambda_2
\left[\frac{5}{28\chi_2}\left(3179-919\chi_2\right.\right.
\nonumber\\
\label{psi_T1}
&&\left.\left. -
228\chi_2^{2}+260\chi_2^{3}\right)x\right]+1\leftrightarrow 2\,.
\end{eqnarray}

We may now use the Love function we introduced in Sec. \ref{sec:love}
as an ``upgrade'' of the concept of Love number and determine new tidal
contributions to the GW signal emitted by an inspiralling binary. Its
$r$-dependency in Eq.~(\ref{lovefit}) may be cast into a dependency on the PN
dimensionless variable $x$ by using the PN expansion of $m/r$ in terms of $x$
\cite{Fav11}. This way one has
\begin{equation}
  k_{2}(x) =\bar{k}_{2}\left[1+\alpha x\right]+\mathcal{O}(x^{2})\,,
\label{k2x}
\end{equation}
and consequently 
\begin{equation}
  \lambda(x) =\bar{\lambda}\left[1+\alpha x\right]+\mathcal{O}(x^{2})\,,
\label{l2x}
\end{equation}
where the coefficient $\alpha$ is determined by means of our dynamical
simulations. The expanded $\lambda_A(x)$'s may now replace the $\bar\lambda_A$'s
in the PN expanded energy ${\cal E}(x)$ and GW flux ${\cal L}_{GW}(x)$,
Eqs.~(\ref{def:Energy}) and (\ref{def:Flux}) respectively. Notice that since the
tidal correction to the binding energy and gravitational flux are known up to
the $1$PN order, Eq.(\ref{k2x}) was truncated at
$\mathcal{O}(m/r)=\mathcal{O}(x)$. We remark that the GW flux depends on the
time derivatives of the quadrupole tensor of the system \cite{VFH11} and,
therefore, the replacement $\bar\lambda_A\rightarrow \lambda_A(x)$ in
Eqns.~(\ref{def:Energy}) and (\ref{def:Flux}) neglects terms arising from time
derivatives of $\lambda_A(x)$. Such terms, however, being proportional to the
velocities of the compact objects, are of a PN order higher than the order of
the expansion in Eq.(\ref{psi_T1}), and may be safely neglected. The replacement
$\bar\lambda_A\rightarrow \lambda_A(x)$ yields the following correction to the
Fourier phase of the gravitational waveform given in Eq.~(\ref{psi_freq}):
\begin{equation}
  \delta\psi_\text{T}(f)=-\frac{3x^{5/2}}{128\nu m^{5}}\bar\lambda_2
  \left[\frac{30}{7}\left(2+27\frac{\chi_1}{\chi_2}\right)
    \alpha_2 x\right]+1\leftrightarrow 2\,,
\label{dpsi_T}
\end{equation}
where $\alpha_A$ are the coefficients appearing in Eq.~(\ref{k2x}) associated to
the $A-th$ body and $\bar\lambda_A$ are determined from
Eq.(\ref{lambdaPP}). Eq.~(\ref{psi_freq}) then becomes
\begin{equation}
\label{psi22}
\psi=\psi_\text{PP}+\psi^\text{N}_\text{T}+\psi^{1\text{PN}}_\text{T}+\delta\psi_\text{T}\,.
\end{equation}
We remark that the term $\psi^{1\text{PN}}_\text{T}$ and our correction
$\delta\psi_\text{T}$ have different origin. The former, obtained independently
in \cite{VFH11} and in \cite{BDF11,DNV12}, is the next-to-leading order tidal
correction to the Fourier phase and was derived under the adiabatic
approximation $Q_{ij}=\bar\lambda C_{ij}$ (or, equivalently, assuming that the
contribution of ``electric'' type quadrupole deformations to the action is
$\Delta S\sim \bar\lambda C_{ij}C_{ij}$); the latter is the result of a
dynamical evolution in which we monitor the variation of the ratio between the
quadrupole and tidal tensor during the inspiral (which we find to be the same,
within $\lesssim 1\%$, for all components $ij$); this variation is encoded in
the parameter $\alpha$ appearing in Eq.~(\ref{l2x}). Thus, these terms (both
increasing as $r$ decreases) simply add linearly.

\begin{figure*}[ht]
\epsfig{file=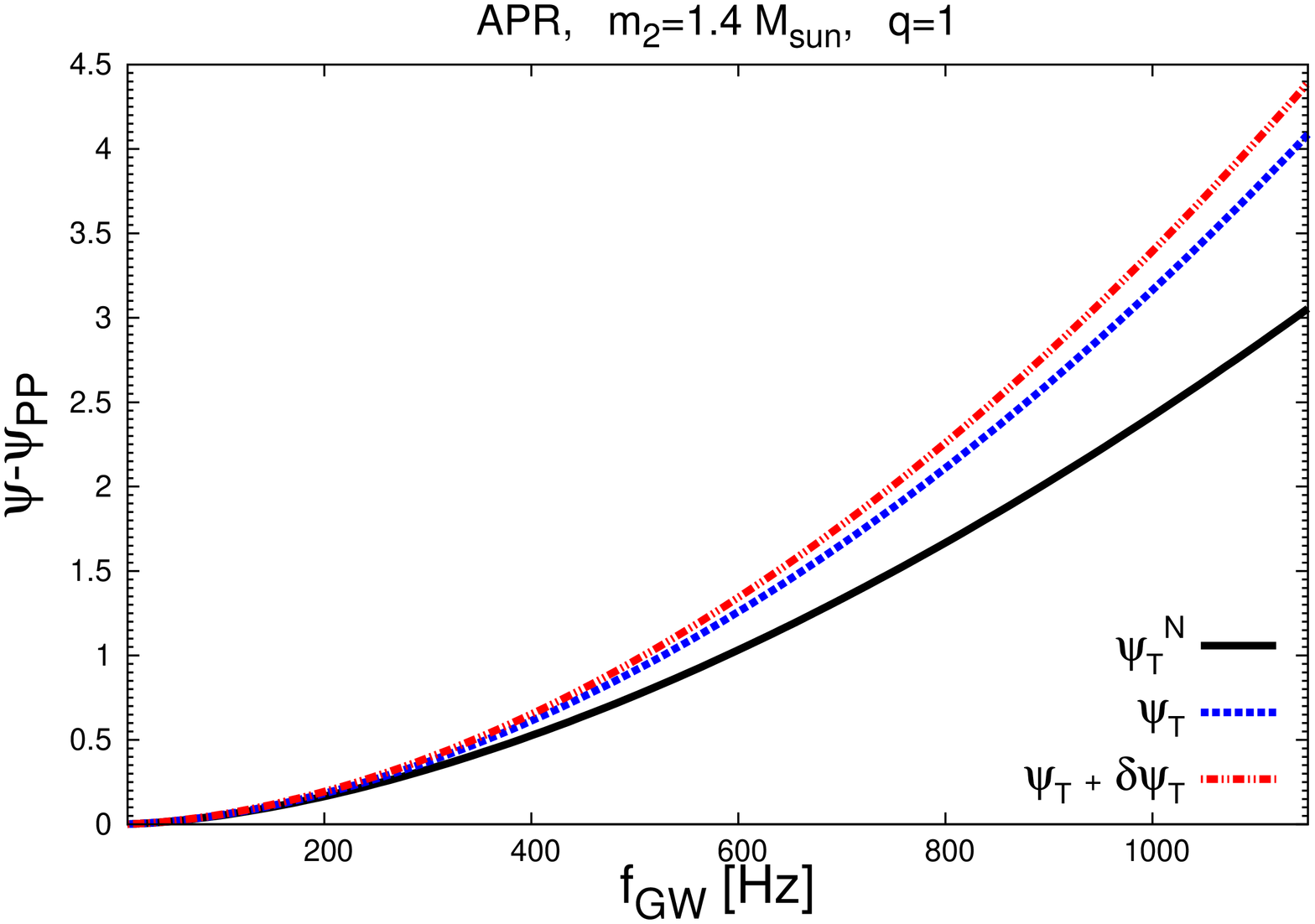,angle=-90,width=250pt}
\epsfig{file=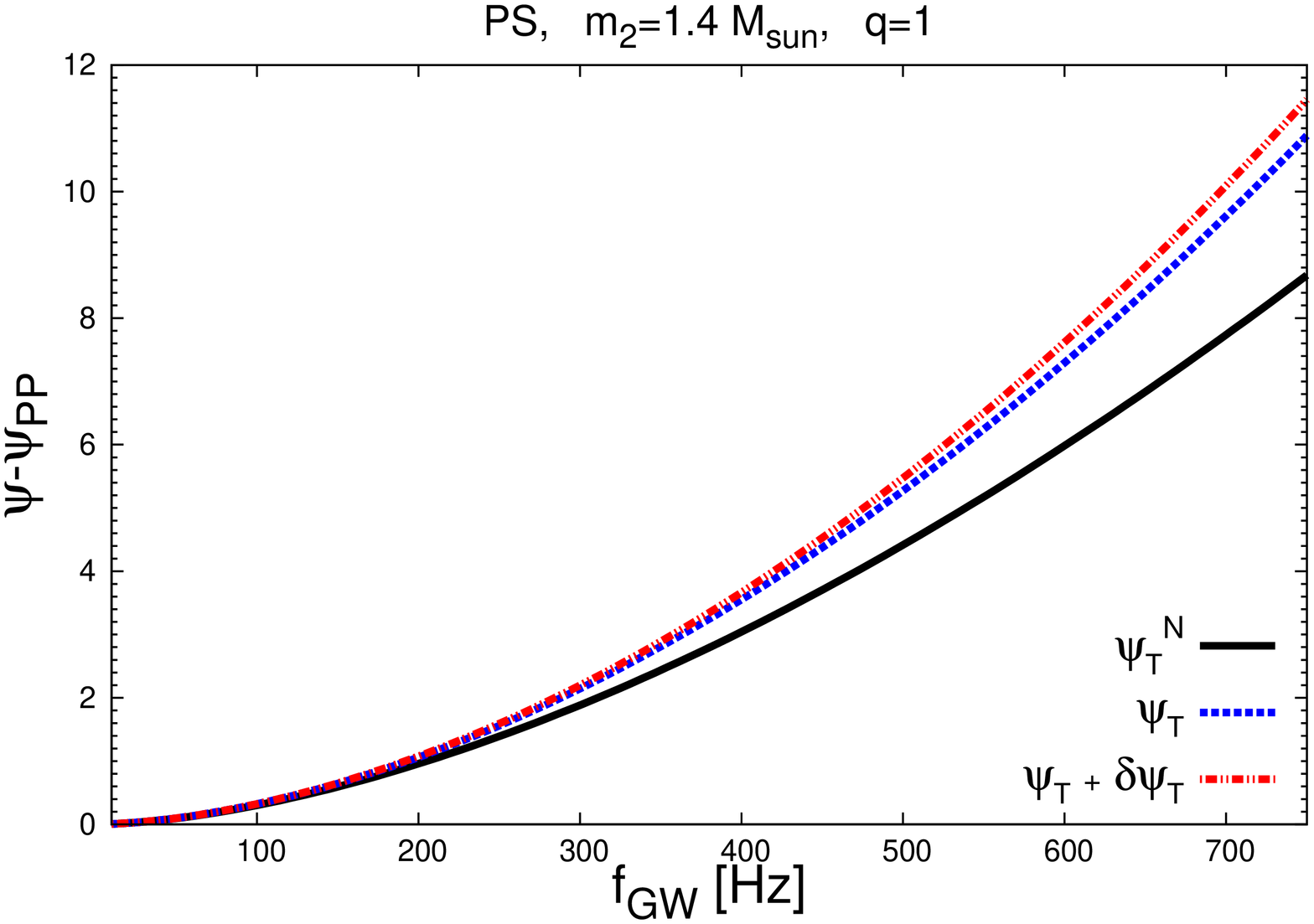,angle=-90,width=250pt}\\
\vspace{1.5cm}
\epsfig{file=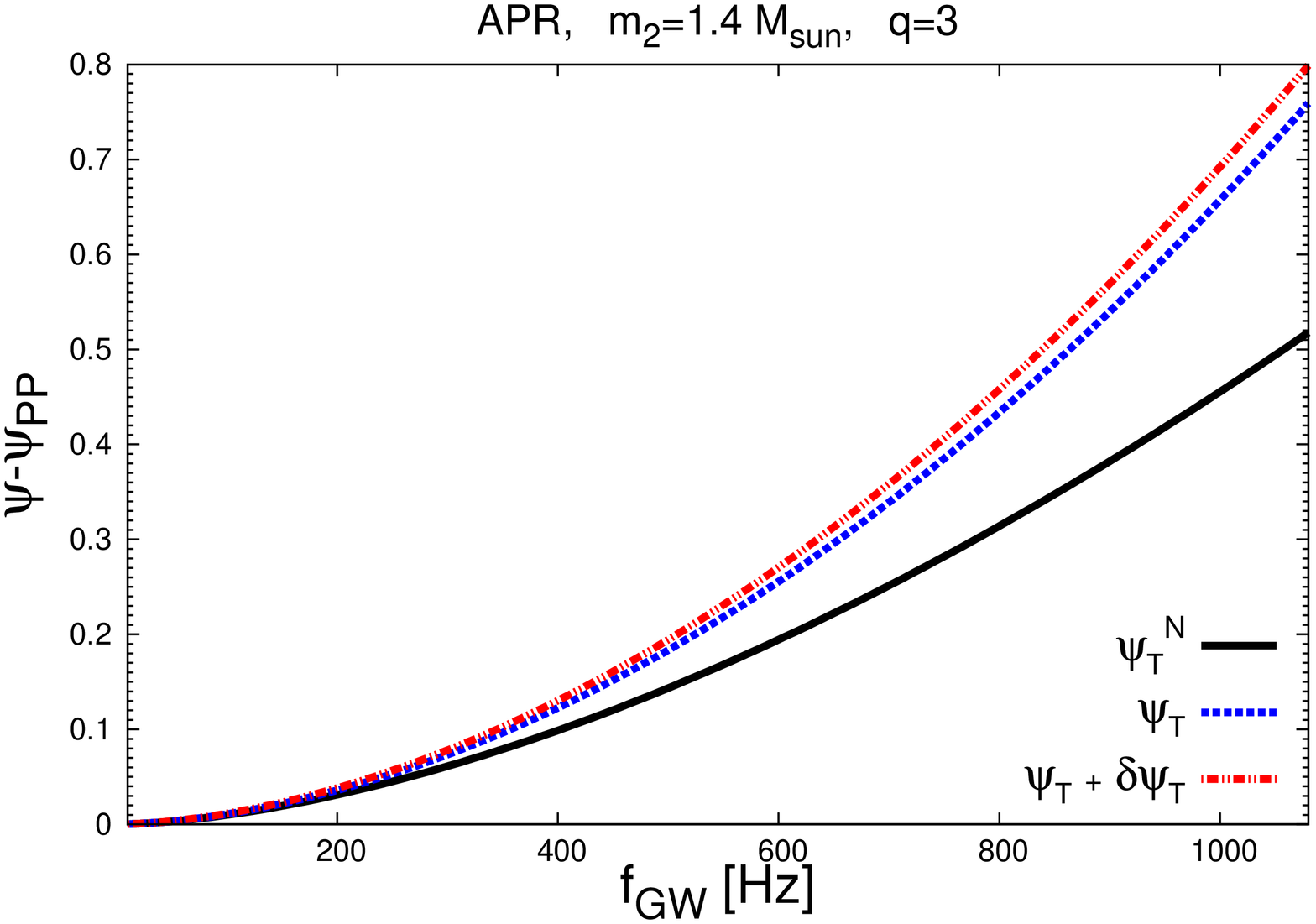,angle=-90,width=250pt}
\epsfig{file=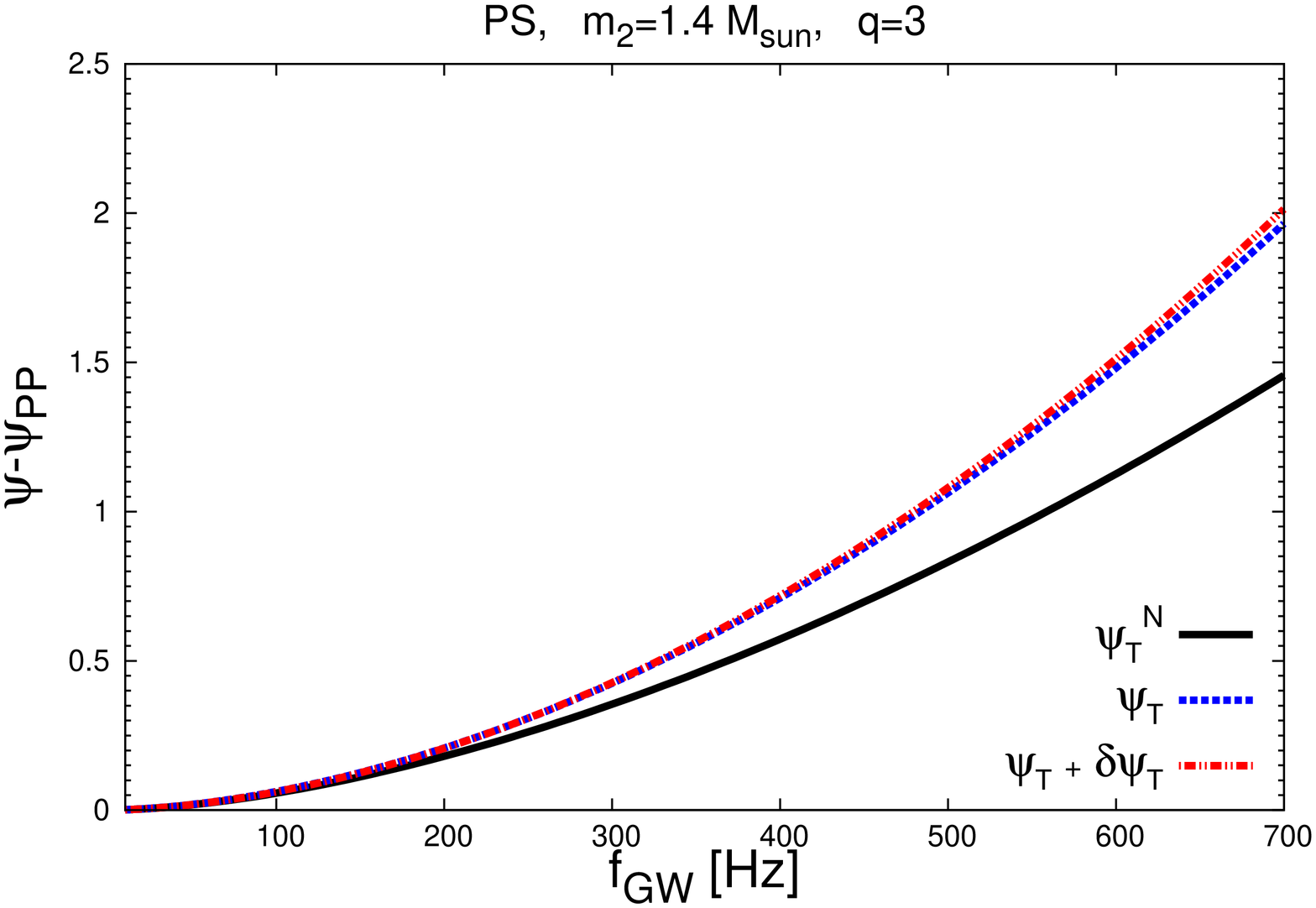,angle=-90,width=250pt}\\
\vspace{1.5cm}
\epsfig{file=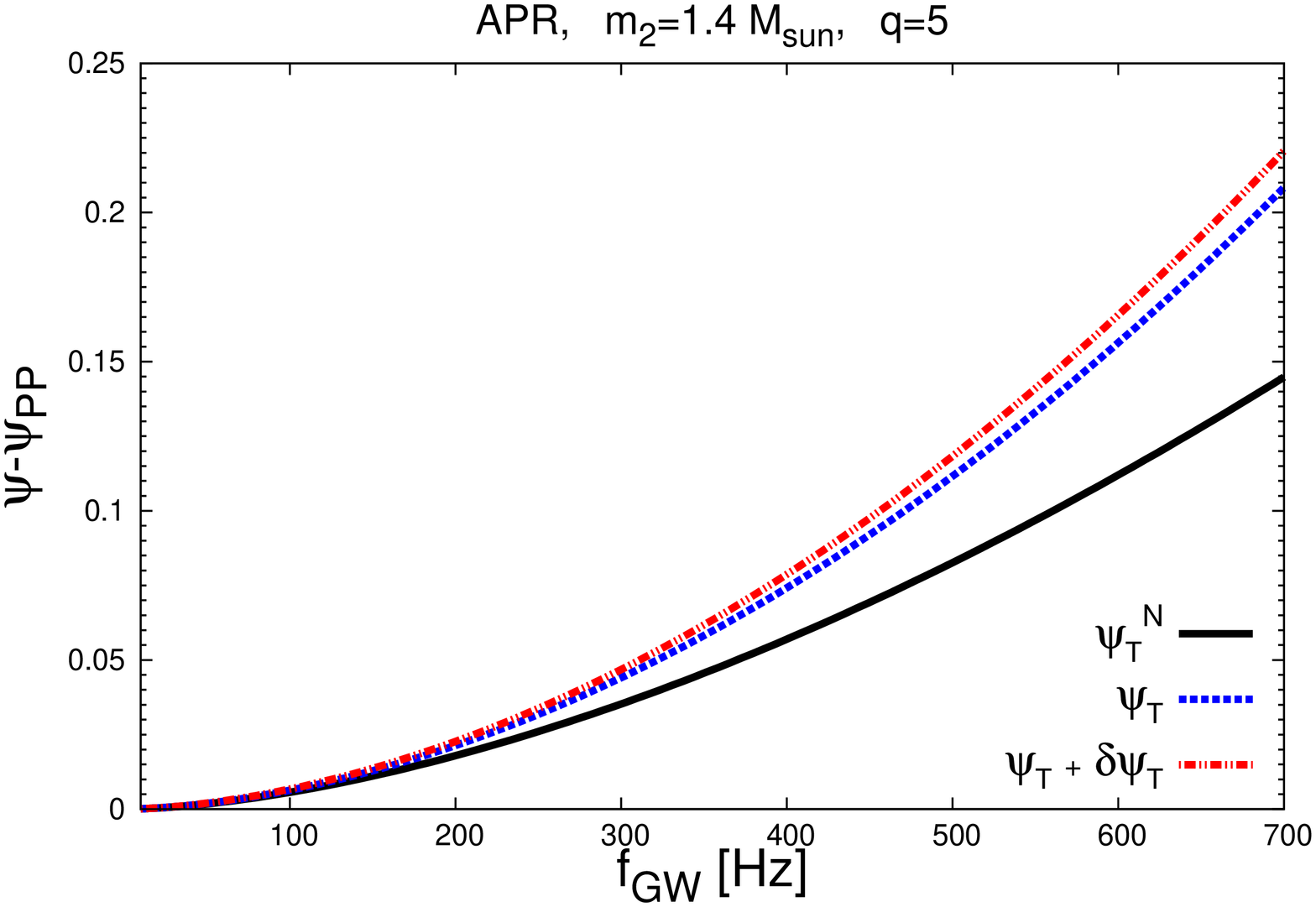,angle=-90,width=250pt}
\epsfig{file=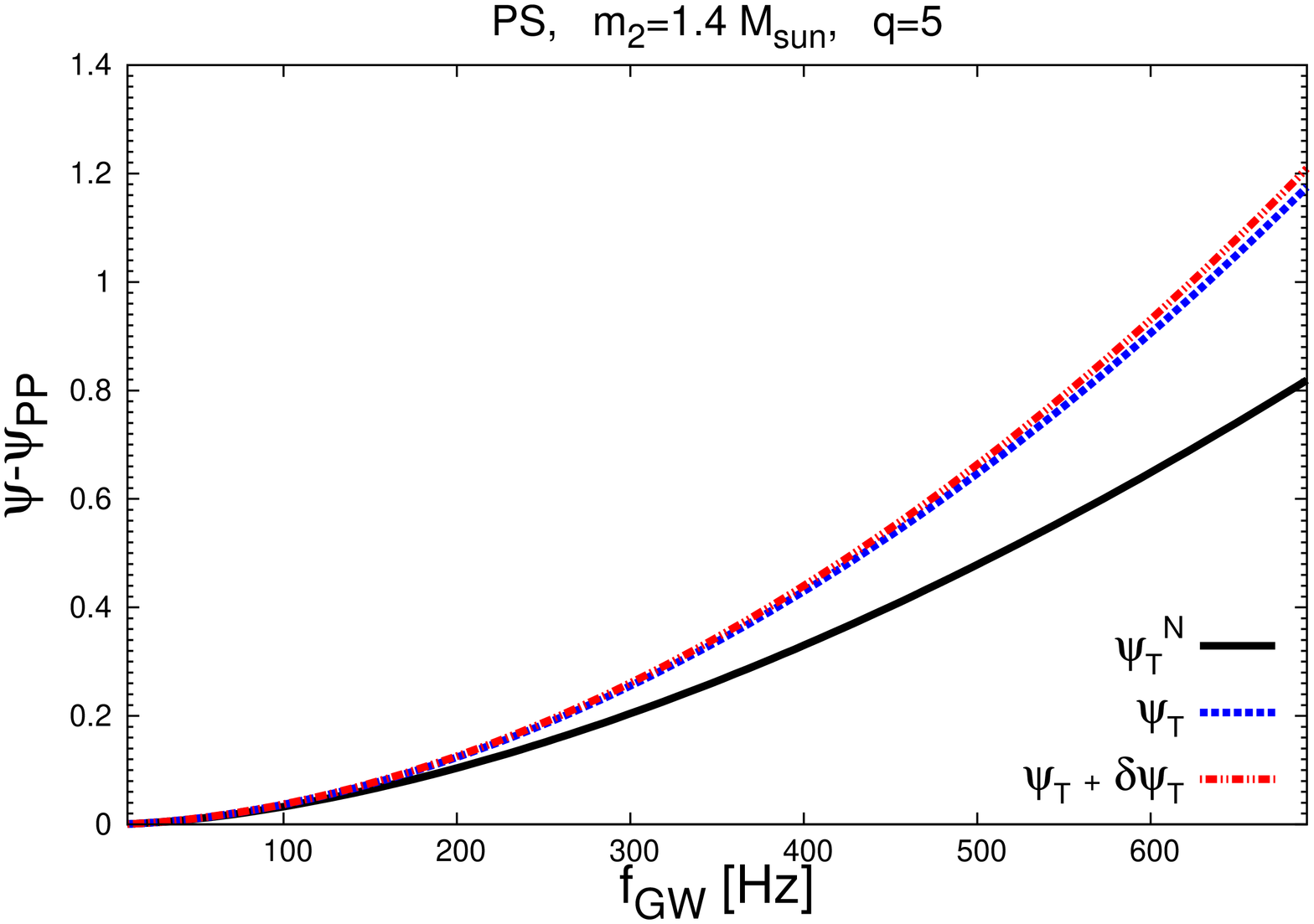,angle=-90,width=250pt}
\caption{(Color online) Tidal contribution to the Fourier phase of the
  gravitational wave signal, for $m_2=1.4\,M_\odot$, the APR/PS EOS
  (left/right panels), and, from top to bottom, $q=1$, $3$, and
  $5$. Different curves show: the leading order term
  $\psi^\text{N}_\text{T}$ (solid black line); the next-to-leading
  order contribution
  $\psi_\text{T}=\psi^\text{N}_\text{T}+\psi^{1\text{PN}}_\text{T}$
  (dotted blue line); the total tidal contribution including our
  correction $\psi_\text{T}+\delta\psi_\text{T}$, (dashed-dotted, red
  line). For each model the phases are shown in the GW frequency range
  $[10\,{\rm Hz},f_\text{cut}]$, where
  $f_\text{cut}=2f^\text{orb}_\text{cut}$, $f^\text{orb}_\text{cut}$
  being the orbital frequencies given in Table
  \ref{tablelove}.}\label{dpsi12}
\end{figure*}

In Fig.~\ref{dpsi12} we compare the different tidal contributions to the GW
Fourier phase: the leading order term $\psi^\text{N}_\text{T}$ (solid black
line); the next-to-leading order
$\psi_\text{T}=\psi^\text{N}_\text{T}+\psi^{1\text{PN}}_\text{T}$ (dotted blue
line); the total tidal contribution including our correction
$\psi_\text{T}+\delta\psi_\text{T}$, (dashed-dotted red line). As discussed
above, for consistency reasons our approach only allows us to include $1/r$
($\sim x$) terms in the gravitational waveform, therefore we truncate our
expansion at order ${\cal O}(x)$.

For each model the phases are shown in the range $[10\,{\rm Hz},f_\text{cut}]$,
where $f_\text{cut}=2f^\text{orb}_\text{cut}$ (see Table
\ref{tablelove}). Fig.~\ref{dpsi12} shows that:
\begin{itemize}
\item The dephasing due to the NS tidal deformation is significantly larger than
  one radiant only for NS-NS binaries. For BH-NS binaries,
  $\psi_\text{T}\lesssim 1$ rad and therefore, as previously established in
  \cite{PROR11}, it is unlikely that second generation detectors such as
  Advanced LIGO/Virgo will detect tidal deformation effects in GWs emitted by
  mixed binaries. Only third generation detectors like the Einstein Telescope
  (ET)~\cite{ET2010}, which could detect signals with very large signal-to-noise
  ratio, would be able to reveal the tidal contribution in BH-NS binary
  coalescences.
\item Looking at systems with $q=1$, i.e., NS-NS binaries, the tidal signal is
  larger for stiffer EOS, or equivalently less compact NSs, and for lower values
  of the NS mass, confirming previous results \cite{HLLR10,DNV12}.
\item The correction derived in this paper, $\delta\psi_\text{T}$, affects the
  GW phase only marginally for the binaries we consider.  Indeed, the two curves
  describing $\psi_\text{T}$ and $\psi_\text{T}+\delta\psi_\text{T}$ nearly
  coincide.
\end{itemize}
Similar conclusions hold when we choose $m_2=1.2M_\odot$.

We remark that in our approach we determine the gravitational waveform up to the
frequency $f_\text{cut}$, corresponding to the onset of the mass-shedding, which
(in our PNA approach \cite{FGM11}) occurs when the NS fills its Roche lobe. As
shown in Table \ref{tablelove}, we find values of $f_\text{cut}$ between
$\sim700$ and $\sim1150$ Hz. These values are larger than the cutoff frequency
assumed in \cite{HLLR10} ($f_\text{cut}=450$Hz), but more conservative than the
values employed in \cite{DNV12}, which correspond to the configuration in which
the surfaces of the two bodies touch. As discussed in \cite{HLLR10}, the main
reason behind the $f_\text{cut}=450$Hz choice is that the next-to-leading order
($1$PN) corrections to the tidal phase, which we include, were neglected.  On
the other hand, the large values for $f_\text{cut}$ used in \cite{DNV12} (see
also \cite{BNTB12}) correspond to the configuration in which the surfaces of the
two bodies touch.  We think that ending the integration when the NS fills its
Roche lobe is a safer choice.

\section{Waveform comparisons}\label{wavecomp}
We now examine the dephasings discussed in the previous section in terms of GW
detection. In order to do so, we calculate \emph{overlaps} and \emph{fitting
  factors} \cite{O96,LOB08} between point-particle waveforms, which we treat as
our templates, and ``real'' signals, i.e. waveforms in which we include tidal
effects to the best of our knowledge by means of Eq.~(\ref{psi22}). Our
templates and signals thus have the form
\begin{eqnarray}
  \label{PP-GW}
  \tilde{h}_\text{PP} &=& \mathcal{A}e^{i\psi_\text{PP}}\\
 \label{realGW}
  \tilde{h}_{\text{T},\delta} &=& \mathcal{A}e^{i(\psi_\text{PP}+
    \psi_\text{T}^\text{N}+\psi_\text{T}^{1\text{PN}}+\delta\psi_\text{T})}\,,
\end{eqnarray}
where $\mathcal{A}$ denotes the GW amplitude  and
$\psi_\text{T}^\text{N}$, $\psi_\text{T}^{1\text{PN}}$, and
$\delta\psi_\text{T}$ are the tidal contributions to the Fourier phase given in
Eqns.~(\ref{psi_T0}), (\ref{psi_T1}), and (\ref{dpsi_T}), respectively.

Given two signals $h_1$ and $h_2$, their noise-weighted inner product is
\begin{equation}
  \langle h_1|h_2\rangle \equiv 4 \Re
  \int_{f_\text{start}}^{f_{\text{cut}}} df \frac{\tilde{h}_1(f)\tilde{h}_2^*(f)}{S_\text{n}(f)}\,,
\end{equation}
where $S_\text{n}(f)$ is the power spectral density of a given detector, the
lower integration bound $f_\text{start}$ depends on the detector one is
examining, and for the upper bound $f_\text{cut}$ we use twice the orbital
frequency $f^\text{orb}_\text{cut}$ discussed in the previous section. More
specifically, we consider second and third generation detectors, such as
Advanced Virgo/LIGO, and the Einstein Telescope, setting $f_\text{start}$ to
$20\,$Hz and $10\,$Hz, respectively, and follow \cite{SS09} for
$S_\text{n}(f)$. The inner product allows us to define the overlap between two
signals, that is, their normalized inner product, maximized over time and phase
shifts:
\begin{equation}
  \mathcal{O}[h_1, h_2] \equiv \max_{\{\tau,\varphi\}}
  \frac{\langle h_1 | h_2 \rangle}{\sqrt{\langle h_1 | h_1 \rangle \langle h_2 | h_2 \rangle}}\,,
\end{equation}
where $\tau$ and $\varphi$ are the time and phase offsets between the two
waveforms. This quantity is useful, for example, when describing in quantitative
terms the effectiveness of a waveform model in detecting a physical
waveform. Furthermore by maximising $\mathcal{O}[h_1, h_2]$ over the intrinsic,
physical parameters of, say, $h_1$, one obtains the fitting factor. In our case,
the point-particle waveform templates, $\tilde{h}_\text{PP}$, depend on the
binary total mass and symmetric mass ratio, so that we denote the fitting factor
with $\mathcal{FF}(m,\nu)$.

As discussed in \cite{LOB08}, when calculating the number of missed events in a
search performed with a discrete template bank, one must take into account the
template bank spacing and the fitting factor $\mathcal{FF}(m,\nu)$ between the
template GW model and the real signal model. In the case of LIGO/Virgo template
banks, this may be done by subtracting a maximum mismatch of $0.03$ to the
fitting factors we calculate: this yields the effective fitting factors
$\mathcal{EFF}(m,\nu)=\mathcal{FF}(m,\nu)-0.03$ which must then be used to
determine the fraction of missed events $1-\mathcal{EFF}(m,\nu)^3$. \cite{LOB08}
additionally discusses a criterion for template waveform accuracy:
$1-\mathcal{FF}(m,\nu)$ must be smaller than $0.005$, so that
$1-\mathcal{EFF}(m,\nu)<0.035$ for LIGO/Virgo.

Our results may be summarized as follows:
\begin{itemize}
\item We find that the lowest overlaps occur in the case of low mass ratios as
  already noted in \cite{PROR11}; this happens for two reasons, in general:
  because low total masses and mass ratios enhance the dephasing in
  Eqs.~(\ref{psi_T0}), (\ref{psi_T1}) and (\ref{dpsi_T}) and because
  $f_\text{cut}$ is higher for lower mass systems, thus allowing the phase
  difference originating from tidal distortions to accumulate over a broader
  frequency range. Furthermore, in the $q=1$ case the phase difference
  originates from the distortion of two NSs instead of a single one.
\item For BH-NS binaries, the $\mathcal{O}[h_\text{PP},h_{\text{T},\delta}]$
  overlaps are $\gtrsim 0.997$ for second generation detectors and $\gtrsim
  0.995$ for the Einstein Telescope, as already found in \cite{PROR11}. This
  means that (1) point-particle templates are within the needed waveform
  accuracy, since $1-\mathcal{FF}\leq 1-\mathcal{O}\leq 0.005$, and that (2) the
  total number of missed events including the effect of template
  bank spacing, is $1-\mathcal{EFF}(m,\nu)^3\leq 1-(\mathcal{O}-0.03)^3\lesssim
  10$\%.
\item For second generation detectors and NS-NS binaries, we find that,
  depending on the EOS, the overlap
  $\mathcal{O}[h_\text{PP},h_{\text{T},\delta}]$ varies within the range
  $[0.963,0.997]$ for $m_2=1.2M_\odot$ and within $[0.982,0.999]$ for
  $m_2=1.4M_\odot$; the lower limits correspond to larger deformabilities,
  i.e. to the stiffer, PS EOS. For third generation detectors, the intervals
  are, instead, $[0.943,0.992]$ and $[0.969,0.997]$.
\item By computing the fitting factors for NS binaries with the PS EOS, we
  obtain $\mathcal{FF}(m,\nu)=0.991$ for $m_2=1.2M_\odot$ and
  $\mathcal{FF}(m,\nu)=0.996$ for $m_2=1.4M_\odot$, in the case of second
  generation detectors. For third generation detectors, one has, instead,
  $\mathcal{FF}(m,\nu)=0.985$ for $m_2=1.2M_\odot$ and
  $\mathcal{FF}(m,\nu)=0.992$ for $m_2=1.4M_\odot$.
\item In the case of Advanced LIGO/Virgo and of a particularly stiff EOS for
  matter in the NS interior, the total number of missed events corresponding
  to the above fitting factors, $1\!-\!\mathcal{EFF}(m,\nu)^3$, may thus be as high
  as $11$\% for low mass NS binaries, whereas it would roughly be $10$\% for a
  canonical $m=2.8M_\odot$ equal mass NS binary. The Einstein Telescope, on the
  other hand, would miss up to $\!\sim\! 13$\% low mass NS-NS inspirals and up to
  $\!11$\%\! canonical NS inspirals. Our results are summarized in Tables
    \ref{Ad_Over}, \ref{ET_Over}.
\item If the EOS of NS matter is very stiff, point-particle templates of binary
  NS inspirals would not meet the required accuracy for Advanced Virgo/LIGO in
  the case of a low total mass, since the fitting factor we obtain yields $1-
  \mathcal{FF}(m,\nu)>0.005$ \cite{LOB08}. The same is true for equal mass NS
  inspirals and the Einstein Telescope, on the high stiffness end of possible
  EOSs. As previously mentioned, the PS EOS is an extreme case of stiff
  EOS. Nevertheless, one should consider this worst (in terms of missed events
  and waveform template accuracy) case scenario and establish if, and eventually
  how, we may remedy; this is important also because a lot of interesting
  physics is associated with low mass NS binary mergers,
  e.g.~\cite{SSKI11,BJHS12}. If one imagines to use templates $\tilde{h}_{T,PN}$
  that include tidal corrections up to (relative) $1$PN order in the Fourier
  phase by means of the \emph{Love Number}, i.e.
\begin{equation}
 \tilde{h}_{\text{T},1\text{PN}} = \mathcal{A}e^{i\psi_\text{PP}+i\psi_\text{T}^\text{N}+i\psi_\text{T}^{1\text{PN}}}
\end{equation}
all overlaps $\mathcal{O}[h_{\text{T,}1\text{PN}},h_{\text{T},\delta}]$ would
differ from unity by less than a part in one-thousand, both for second and third
generation detectors. To the best of our knowledge in the modelling of the real
signal, the inclusion of Love Number dependent tidal terms in GW templates would
therefore greatly help in constraining the NS EOS in post-processing analysis,
confirming the result of \cite{DNV12}. Building gravitational waveforms within
the adiabatic approximation (i.e., including $\psi_\text{T}^\text{N}$ and
$\psi_\text{T}^{1\text{PN}}$) is thus very reliable.
\item Let us now suppose that the EOS of NS matter is stiff and that an
  $m=2.4M_\odot$ equal mass binary NS inspirals close enough to Earth. The
  detection of its gravitational radiation by a second generation detector with
  an imaginarily infinitesimally spaced template bank would hit a point-particle
  template with total mass $m=2.446M_\odot$ and symmetric mass ratio
  $\nu=0.242$, since these are the values that maximise the overlap
  $\mathcal{O}[h_\text{PP},h_{\text{T},\delta}]$: this means that tidal effects
  contribute with a $2$\% error to the total mass measurement and with a $3$\%
  error on the symmetric mass ratio measurement. In the case of an
  $m=2.8M_\odot$ equal mass inspiral, the tidal contributions to the errors on
  the total mass and the symmetric mass ratio measurements would be $1$\% and
  $2$\%, respectively. The same values hold for third generation detectors.
\end{itemize}

\begin{table}[ht]
  \centering
  \begin{tabular}{cccccccc}
    \hline
    \hline
    EOS & $m(M_{\odot})$ && ${\cal O}[h_{PP},h_{T,\delta}]$ && $\mathcal{FF}(m,\nu)$ &&  $\#$ missed \\  
     \hline
    APR & $2.4$ && $0.997$ && $1.000$ && $0$\% \\
        & $2.8$ && $0.999$ && $1.000$ && $0$\% \\
     PS & $2.4$ && $0.963$ && $0.991$ && $11 \%$\\
        & $2.8$ && $0.982$ && $0.996$ && $9.8\%$\\
    \hline
    \hline
  \end{tabular} \caption{We report the values of  overlaps  ${\cal O}[h_{PP},h_{T,\delta}]$ and fitting factors 
  $\mathcal{FF}(m,\nu)$ between point-particle templates and waveforms in which we include tidal effects 
  evaluated as in Eq.~(\ref{psi22}). We also show the total number of missing events. All values refer to 
  NS-NS systems and second generation detectors AdVirgo/LIGO. \label{Ad_Over}}
  \end{table}

\begin{table}[ht]
  \centering
  \begin{tabular}{cccccccc}
   \hline
    \hline
    EOS & $m(M_{\odot})$ && ${\cal O}[h_{PP},h_{T,\delta}]$ && $\mathcal{FF}(m,\nu)$ &&  $\#$ missed\\  
     \hline
    APR & $2.4$ && $0.992$ && $1.000$ && $0$\% \\
        & $2.8$ && $0.997$ && $1.000$ && $0$\% \\
     PS & $2.4$ && $0.943$ && $0.985$&& $13 \%$\\
        & $2.8$ && $0.969$ && $0.992$&& $11 \%$\\
    \hline
    \hline
  \end{tabular} \caption{The same quantites of Table \ref{Ad_Over}, but for third generation detector ET. \label{ET_Over}}
  \end{table}

\section{Concluding remarks}\label{conclusions}
In this paper we improve the PNA model recently developed in \cite{FGM11}, by
computing the tidal tensor of a binary system up to $2$ PN order.  Using this
approach, we study the dynamical evolution of NS deformations in compact binary
inspirals, finding a number of interesting results.

We find that the ratio between the quadrupole tensor $q_{ij}$ and the tidal
tensor $c_{ij}$ increases during the last stages of the inspiral, and we show that
the Love number has to be considered as the asymptotic value of this ratio for
sufficiently large orbital separation.  This increase, however, only marginally
affects the phase of the emitted signal, thus assessing the validity of the
adiabatic approximation when modeling the gravitational wave signal.

We provide a semi-analytical expression of the Love number, in terms of the
scalar quadrupole moment and of the integral of pressure over the stellar
volume, both computed for the star at isolation, i.e.  for a spherically
symmetric configuration (Eq.~(\ref{k2an})). This expression provides a physical
insight on the Love number and a formulation alternative to that given in
\cite{H08}, only in terms of quantities which refer to the structure of the star
in its spherical equilibrium configuration.

In addition, we estimate the reliability of point particle templates with
respect to a ``real'' signal modeled by means of our approach, finding that such
templates marginally fail to meet the standard accuracy requirements
($1-\mathcal{FF}(m,\nu)<0.005)$ for NS-NS inspirals with very stiff equation of
state.

\section*{Acknowledgements}
We thank L.~Rezzolla, A.~Nagar, and D.~Bini for useful discussions. This work
was partially supported by CompStar, a research networking program of
the European Science Foundation. F.P. was supported in part by the DFG grant
SFB/Transregio~7.


\end{document}